\documentclass[fleqn,usenatbib]{mnras}

\pdfoutput=1    
\usepackage{newtxtext,newtxmath}
\usepackage{times}

\usepackage[T1]{fontenc}
\usepackage{ae,aecompl}

\usepackage{graphicx}	
\usepackage{amsmath}	
\usepackage{amssymb}	
\usepackage[caption = false]{subfig}
\usepackage{graphics}	
\def\aj{AJ}%
\def\araa{ARA\&A}%
\def\apj{ApJ}%
\def\apjl{ApJ}%
\def\apjs{ApJS}%
\def\ao{Appl.~Opt.}%
\def\apss{Ap\&SS}%
\def\aap{A\&A}%
\def\aapr{A\&A~Rev.}%
\def\aaps{A\&AS}%
\def\mnras{MNRAS}%
\def\pasa{PASA}%
\def\pasp{PASP}%
\def\solphys{Sol.~Phys.}%
\def\nat{Nature}%
\def\iaucirc{IAU~Circ.}%
\def\jgr{J.~Geophys.~Res.}%
\def\planss{Planet.~Space~Sci.}%
\newcommand{\fermi}{{\it Fermi}}
\DeclareRobustCommand{\VAN}[3]{#2}

\title[Properties of low frequency compact sources]{Interplanetary Scintillation studies with the Murchison Wide-field Array II: Properties of sub-arcsecond compact sources at low radio frequencies}

\author[R. Chhetri et al.]
{\parbox{\textwidth}{R.~Chhetri,$^{1,2}$\thanks{E-mail: \texttt{rzn.chhetri@gmail.com}}
J.~Morgan,$^{1}$
R.~D.~Ekers,$^{1,3}$
J-P~Macquart,$^{1,2}$
E.~M.~Sadler,$^{2,4}$
M.~Giroletti,$^{5}$
J.~R.~Callingham$^{6}$ and
S.~J.~Tingay$^{1}$}\vspace{0.4cm}\\
\parbox{\textwidth}{$^{1}$International Centre for Radio Astronomy Research, Curtin University, GPO Box U1987, Perth, WA 6845, Australia\\
$^{2}$ ARC Centre of Excellence for All-Sky Astrophysics (CAASTRO), Australia \\
$^{3}$CSIRO Astronomy and Space Science (CASS), Marsfield, NSW 2122, Australia\\
$^{4}$Sydney Institute for Astronomy, School of Physics A28, The University of Sydney, NSW 2006, Australia \\
$^{5}$INAF Instituto di Radioastronomia, via Gobetti 101, Bologna I-40129, Italy\\
$^{6}$Netherlands Institute for Radio Astronomy (ASTRON), PO Box 2, 7990 AA Dwingeloo, The Netherlands}}

\date{Accepted XXX. Received YYY; in original form ZZZ}

\pubyear{2017}

\begin{document}
\label{firstpage}
\pagerange{\pageref{firstpage}--\pageref{lastpage}}
\maketitle

\begin{abstract}
We report the first astrophysical application of the technique of wide-field Interplanetary Scintillation (IPS) with the Murchison Widefield Array (MWA). This powerful technique allows us to identify and measure sub-arcsecond compact components in low-frequency radio sources across large areas of sky without the need for long-baseline interferometry or ionospheric calibration. We present the results of a five-minute observation of a 30$\times$30\,deg$^2$\ MWA field at 162\,MHz with 0.5 second time resolution. Of the 2550 continuum sources detected in this field, 302 (12 per cent) show rapid fluctuations caused by IPS. 
We find that at least 32\% of bright low-frequency radio sources contain a sub-arcsec compact component that contributes over 40\% of the total flux density. 
Perhaps surprisingly, peaked-spectrum radio sources are 
the dominant population among the strongly-scintillating, low-frequency sources in our sample. While gamma-ray AGN are generally compact, flat-spectrum radio sources at higher frequencies, the 162\,MHz properties of many of the \fermi\ blazars in our field are consistent with a compact component embedded within more extended low-frequency emission. 
The detection of a known pulsar in our field shows that the wide-field IPS technique is at the threshold of sensitivity needed to detect new pulsars using image plane analysis, and scaling the current MWA sensitivity to that expected for SKA-low implies that large IPS-based pulsar searches will be feasible with SKA. Calibration strategies for the SKA require a better knowledge of the space density of compact sources at low radio frequencies, which IPS observations can now provide. 
\end{abstract}




\section{Introduction}

Since the 1950s, the radio sky has been mapped by a series of surveys, with increasing depth and over an increasing range of frequencies (see Figure 1)\footnote{See \citet{Callingham:phdthesis} for an historical overview.}.
Most recently, a new generation of radio telescopes including LOw Frequency ARray \citep[LOFAR;][]{VanHaarlem2013} and the Murchison Widefield Array  \citep[MWA;][]{Tingay2013}, as well as upgrades to existing facilities, has led to a plethora of low-frequency ``all-sky'' radio surveys.
These include MSSS \citep{Heald2015} with LOFAR, GLEAM with the MWA \citep{Hurley-Walker2017}, TGSS with the GMRT \citep{Intema2017}, and VLSSr \citep{Lane2014}, a redux of VLSS \citep{Cohen2007},  with the VLA.

High-resolution observations play an important role in determining the astrophysical nature of individual radio sources.
However surveying large number of sources with the long baselines required to achieve high resolution is technically challenging, especially at frequencies below 1\,GHz.
Early attempts used a single sensitive baseline \citep[e.g.][]{Porcas2004}, while more recent attempts used a multi-phase centre approach \citep[e.g.][]{Middelberg2011}.
Such surveys have never matched more conventional surveys in terms of numbers of sources or survey area.
By far the most impressive has been that of \citet{Deller2014}, but even with 408 hours of observing time, and the use of multiple phase centres, less than 3\% of the sources in the FIRST survey \citep{Becker1995} have been surveyed.

At lower frequencies, the wide fields of view, which make surveys so efficient at these wavelengths, may be utilized \citep{Lenc2008}.
However, this may require solving for an independent ionosphere for every interferometer element, for every source.
This is possible provided that the sources themselves are sufficiently strong to act as calibrators \citep{Moldon2015,Jackson2016}.
Nonetheless, to date, the only ``all-sky'' survey of sub-arcsecond radio emission has been carried out at 20\,GHz, where a 6\,km baseline is sufficient to resolve 0.15\arcsec\ scales \citep{Chhetri2013}.

In previous decades the Interplanetary Scintillation (IPS) phenomenon discovered by \citet{Clarke:phdthesis} was used to determine which sources contained sub-arcsecond structure \citep{Hewish1964}.
The MWA provides excellent instantaneous UV coverage and time resolution sufficient to measure IPS.
\citet[][hereafter Paper I]{Morgan2017} have recently demonstrated that IPS can be measured for many hundreds of sources across the entire 900\,deg$^2$\ MWA field of view (at $\sim$150 MHz), making it possible to identify which radio sources have compact components at low frequencies and measure their flux densities on sub-arcsecond scales, using just a few minutes of data.

In this paper, the second in a series, we present the first study of the astrophysical properties of sub-arcsecond compact sources selected at 162 MHz using the widefield IPS technique.
We have chosen a new field which has full overlap with both the GLEAM survey (which provides continuous spectra from 72-231 MHz) and the AT20G survey (the only complete 20 GHz survey of the sky, and the only large radio continuum survey for which the source compactness on sub-arcsecond scale is known for every source).

As in Paper I, the data analysed in this paper come from a single 5-minute MWA observation with 0.5\,s time resolution, with the Sun just beyond the field of view.
Measurements of IPS then allow us to characterise the sub-arcsecond compactness of all sources detected within the $\sim$900\,deg$^2$ field of view of the MWA.
Section~\ref{sec:obs} of this paper describes the observations and the calibration, imaging, and identification of continuum sources.
Section~\ref{sec:identification} describes the method used to identify and measure IPS sources, and also presents the catalogue.
The general properties of sources showing IPS are discussed in Section~\ref{sec:discussion}, while Section~\ref{sec:interesting} describes a few individually-interesting sources in more detail.
Finally, we summarize our main findings in Section~\ref{sec:summary}. 

Throughout this paper we have assumed a lambda CDM cosmology with $\Omega_m$ = 0.27, $\Omega_{\lambda}$ = 0.73, and $H_0$ = 67.8 $\rm km\, s^{-1} Mpc^{-1}$.
For radio spectral-index measurements, we use the definition $S_{\nu} \propto \nu^{\alpha}$ where S, $\nu$, and $\alpha$ are flux density, frequency, and spectral index respectively.

\begin{center}
\begin{figure*}
\includegraphics[scale=0.4, angle=0]{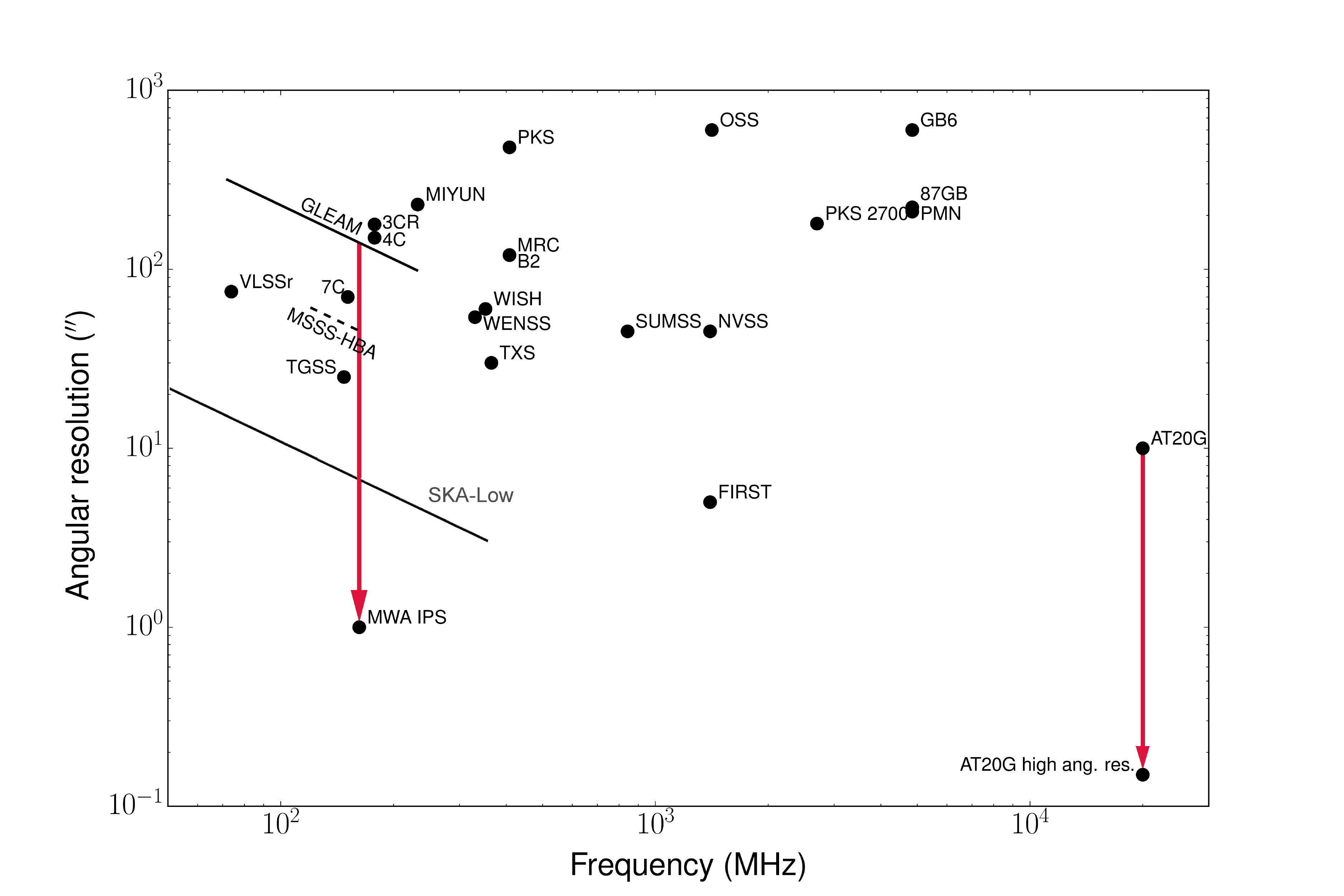}
\caption{A comparison of frequency coverage and their angular resolutions for some widefield large-area radio surveys, covering $>$ 1 steridian of the sky. The point marked ``AT20G high ang. res. catalogue" shows angular resolution of the AT20G high-angular-resolution catalogue \protect{\citep{Chhetri2013}}, and ``MWA IPS" is the lower limit on angular resolution from the current paper. Excellent historical overview of surveys (including ones shown in this figure) and their parameters are presented in \protect{\citet{Callingham:phdthesis}}. Angular resolution for the future SKA-low, based on current specifications of 65\,km longest baselines, is also presented for comparison.}
\label{Fig:SurveysAngRes}
\end{figure*}
\end{center}

\section{Observations and initial data reduction}
\label{sec:obs}
The data used in this study was taken by the Murchison Widefield Array, located in Western Australia, and consists of 2048 dual-polarization dipole antennas. These antennas are arranged into 128 ``tiles'' of 4 $\times$ 4 dipoles, distributed across an area roughly 3\,km in diameter. This arrangement of a large number of small elements results in an interferometer with outstanding instantaneous UV coverage coupled with an exceptionally wide field of view \citep{Lonsdale2009}.

The work presented here uses a single 286\,s ($\approx$5 minutes) observation made on 1 May 2016 at 02:25:20 UTC (during the daytime), with the Sun placed in the first null of the primary beam. The pointing centre was 44 degrees from the Sun (the minimum and maximum solar elongations for sources detected in the field are 20.7 and 69.6 degrees respectively) at RA 00:49:02 and Dec -19:58:54, and close to the south Galactic pole. We split the MWA's instantaneous bandwidth into two 15.4\,MHz bands. Only the upper half, centred at 162\,MHz, is discussed here. The data were correlated in real-time on site, using the standard MWA correlator \citep[]{Ord2015} with the maximum available time resolution of 0.5\,s.

The parameters of this observation are almost identical to that studied in Paper I, except that the field is slightly further from the Sun, and the observing frequency has been shifted slightly to match the GLEAM survey. We followed almost exactly the same calibration and data reduction approach with one exception: since the ionospheric conditions were far more settled during this observation, accounting for image distortion was much simpler. Furthermore with a uniform ionosphere across the field of view, we were able to successfully self-calibrate the field to achieve far higher image fidelity.

\subsection{Calibration and Imaging}
\label{sec:calibration}
The ionosphere would normally be a major issue for any sub-arcsecond imaging over a wide field of view at these frequencies but IPS observations can be conducted using only baselines short enough to eliminate ionospheric decorrelation or non-isoplanicity in all but the most extreme ionospheric conditions. 
Information about sub-arcsecond structure is encoded in the IPS and cannot be removed by the ionosphere. For our observation, the shortest useful baseline at which unresolved source confusion becomes a limiting factor is $\sim$1200\,m (approximately one third of the longest baseline of the MWA used for this observation).

A pre-dawn observation of the primary calibrator Hydra A was used to provide an initial calibration sufficient to make a synthesis image of the field using the full 5 minutes of data. WSClean and associated tools \mbox{\citep{Offringa2014,Offringa2016}} were used for calibration and imaging throughout. A single round of self-calibration was then carried out, closely following the method outlined in \citet{Hurley-Walker2017}, to produced a single Stokes I image again using the $\sim$5 minute data. Following Paper I we term this the ``standard image''. We then used the same solution to produce XX and YY images for each of 584 separate half-second intervals. These images form the basis of our time series analysis.

\subsection{Source finding and cross-matching}
\label{sec:sourcefinding}
The Aegean source finding tool \citep{Hancock2012} was then used on the standard image to identify all of the sources, scintillating and non-scintillating, in our field. No primary beam correction was applied to maintain a relatively uniform noise level. 2567 sources were identified in our field by Aegean using default parameters.

In paper I, careful correction of ionospheric distortions was necessary in order to unambiguously find counterparts of scintillating sources in TGSS.
This is not necessary in this case since we are not cross-matching with TGSS, but with GLEAM, which has almost an order of magnitude lower resolution (see Figure~\ref{Fig:SurveysAngRes}), thus even with cross-matching errors of an arcminute or more, the correct GLEAM counterpart can be identified unambiguously.

Furthermore, a preliminary cross-match with GLEAM revealed that the ionospheric shifts were almost perfectly uniform across the field of view, and the ionospheric effects could be accounted for by simply shifting all source positions by the same vector (approximately 55\,arcsec to the West).
After this correction was applied, the average offset dropped to $\sim$10\,arcsec, or about 10\% of the width of the synthesised beam -- the best achievable astrometric precision given our minimum S/N of 5 \citep{Reid1988} in our Stokes I image.

We then cross-matched these object positions with the GLEAM catalogue positions.
The MWA GLEAM survey \citep{Wayth2015} covers all declinations below +30\degr, with a total sky coverage of over 30\,000 square degrees. The GLEAM catalogue \citep{Hurley-Walker2017} provides flux densities from 74 to 231 MHz in 20 equally-spaced frequency bands. Source detection was carried out using a wide band centred on 200\,MHz (the optimal trade off between maximising sensitivity and minimising confusion noise) where the image RMS was typically 10\,mJy and the angular resolution was $\sim$ 2\arcmin. For our field and frequency range, the typical GLEAM resolution and RMS were 140\arcsec\ and 23\,mJy respectively (in the 166\,MHz GLEAM band). We used TOPCAT \citep{Taylor2005} with a search radius of 1.5 arcminutes to make a cross-match between our standard image and the GLEAM catalogue. 2550 of the 2567 sources with S/N ratio $\geq$ 5 in our standard image have GLEAM counterparts\footnote{The 17 sources without GLEAM counterparts do not have TGSS or NVSS counterparts and many are clearly sidelobes or close to the edge of the image. There is no indication that any are real detections.}. These sources form the sample that we study in this paper.

\section{Identifying and characterising Scintillating Sources}
\label{sec:identification}
We next converted our set of individual time step images (see Section~\ref{sec:calibration}) into the image cube format described in Paper I.
This facilitated easy extraction of the time series of Stokes I flux densities corresponding to any pixel within our image.
The brightest pixel of each of our sources could therefore easily be examined for the temporal signature we expect from IPS, either by direct examination of the time series or by generating the autocorrelation function or power spectrum. For each source, an additional four pixels at four corners offset by 7 pixels were also extracted. These offset pixels should show no scintillation unless an unrelated source or the sidelobe of a strong scintillating source falls on the off-source pixel. 

\begin{center}
\begin{figure*}
\includegraphics[scale=0.550, angle=0]{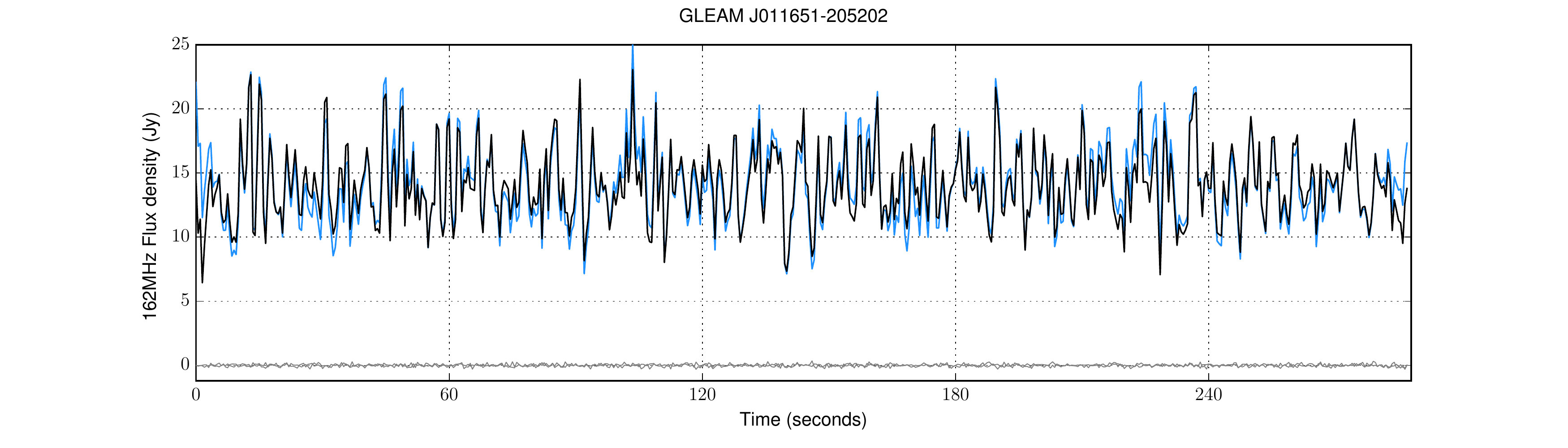}
\includegraphics[scale=0.550, angle=0]{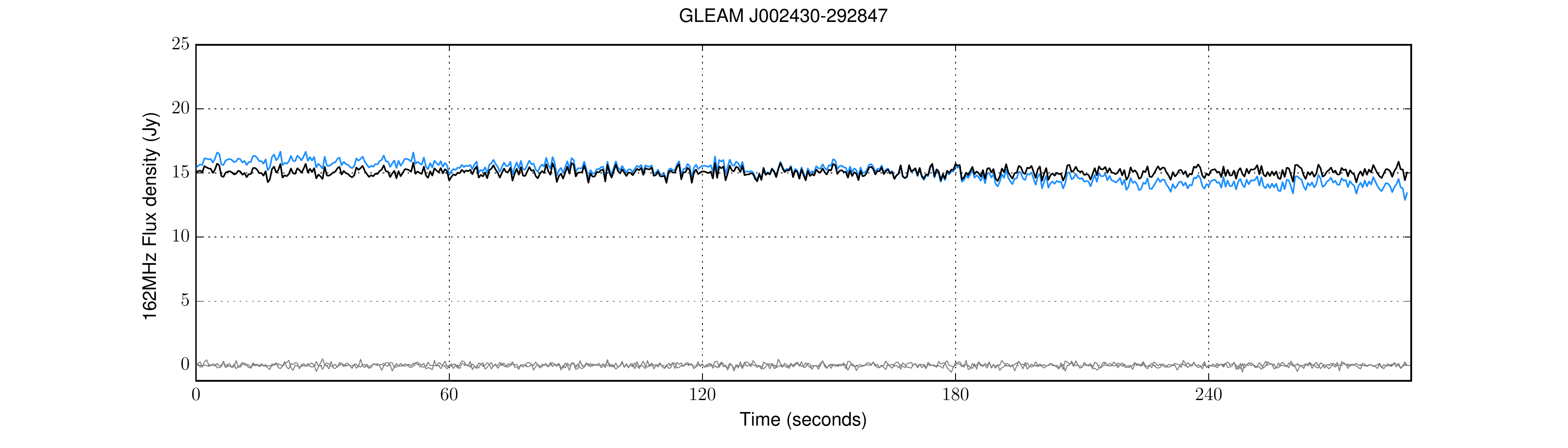}
\caption{Time series plot for a strongly scintillating source (top) and a source which shows little or no scintillation (bottom). The measured time series of the source is shown in blue. The black line represents the time series after applying a high pass filter to remove any effects of the ionosphere or instrument. Time series for two off-source pixels representing noise values are plotted at the bottom in grey colour.}
\label{Fig:timeSeries}
\end{figure*}
\end{center}

In Figure \ref{Fig:timeSeries}, we present examples of time series for representative bright sources, one showing clear scintillation and another with little or no IPS. In a small number of cases time series of the source (shown in blue in Figure \ref{Fig:timeSeries}) also shows lower frequency variations, as seen in the case of non-scintillating source in Figure \ref{Fig:timeSeries}. Such lower frequency variations with periods $\geq$10 seconds, lower than IPS frequencies, may be introduced by ionosphere or instrumental effects. Application of a high pass filter to our data removes these variations effectively (as shown in Figure \ref{Fig:timeSeries}). All analyses presented in this paper were made using the filtered data. 

In order to identify all scintillating sources within our field, and to derive scintillating flux densities and their uncertainties (and upper limits for those sources where any scintillation does not reach the detection threshold), we use the variability imaging technique.
Below we summarise this technique with particular emphasis on those aspects of our procedure which differ from the approach taken in Paper I.
The reader is referred to the previous paper for a full description and validation of the variability imaging technique.

\subsection{Variability image}
\label{sec:var_image}
\begin{center}
\begin{figure*}
\includegraphics[scale=0.93, angle=0]{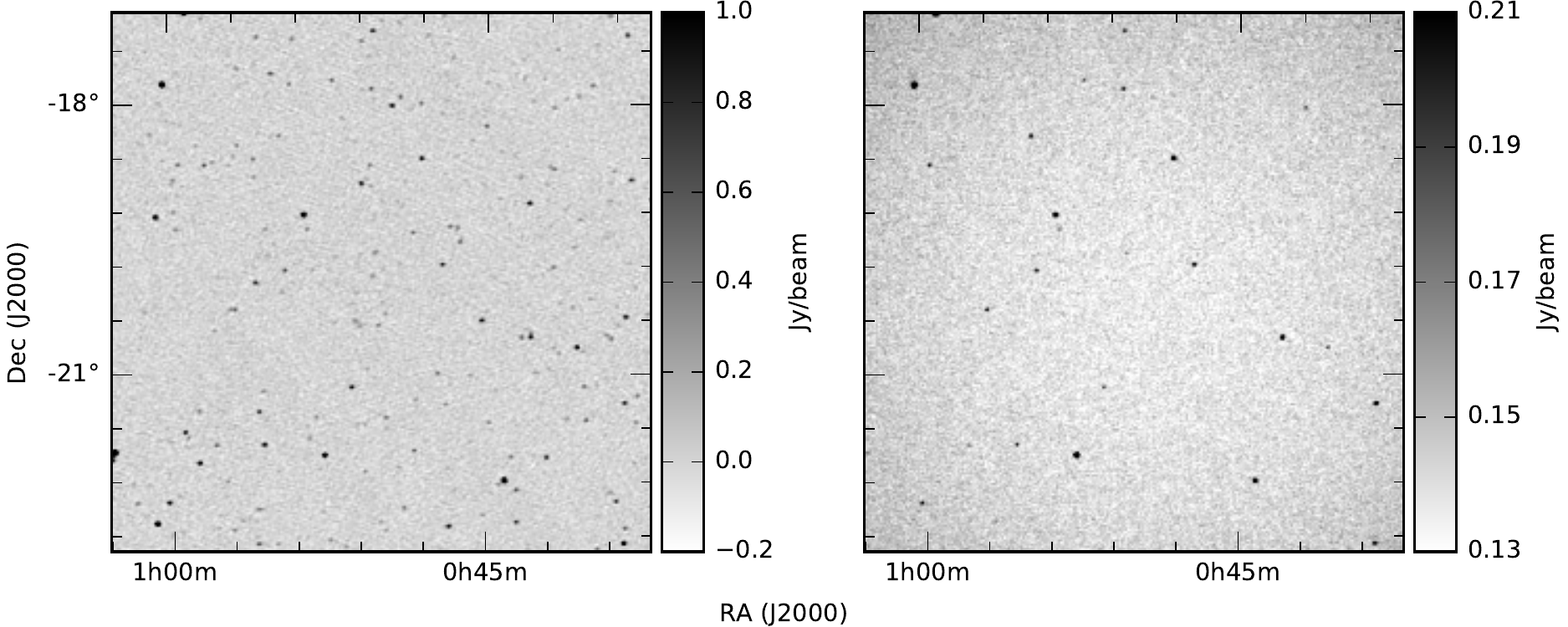}
\caption{These two images show the effectiveness of the variability image in identifying scintillating sources. The image on the left is a small part of the standard image towards the centre of our observation field. The image on the right shows only the scintillating sources in the same area of the sky.}
\label{Fig:SDvsIimg}
\end{figure*}
\end{center}

The variability image (introduced in Paper 1) consists of a 2D image where each pixel value represents the standard deviation of the time series after apply a filtering to remove timescales which may be corrupted by ionospheric activity \footnote{Specifically, we used a Butterworth high-pass filter of order 2 with cut-off frequency of 0.05Hz.}.
Thus the only variance that remains is that due to IPS and system noise. Figure \ref{Fig:SDvsIimg} shows a small portion of our variability image along with a corresponding section of the standard image.  This illustrates clearly how effective this technique is at filtering out extended emission such as that emanating from the diffuse lobes of radio galaxies, leaving only that emitted from sub-arcsecond structures indicated by IPS. 
A convenient property of the variability image is that in the signal-absent case, the statistics of the pixel values $P$, are very close to a Gaussian where the mean ($\mu$) is the system noise in a single timestep. As in a standard noisy image, the standard deviation of the pixel values, $\sigma$, is a useful statistic for measuring the noise level in the image. These properties mean that we can use a source finding program (such as Aegean) to find the variable sources where the signal ($P-\mu$) exceeds a chosen level of significance. We ran Aegean to identify sources that have S/N ($\left(P-\mu\right)/\sigma$) of 4 or higher, generating a list of 952 potential scintillators. 
The S/N of 4 (rather than 5 as in paper I) was chosen since, rather than doing a blind search for IPS sources, we wish to measure the scintillating flux densities of all of the sources in the catalogue we have defined based on known GLEAM counterparts  (Section~\ref{sec:sourcefinding}).  

Cross-matching the variability image detections against our catalogue yielded 302 matches.
Of the variability image detections which did \emph{not} have a match, all but one had a S/N below 5.
The only exception was a single 5.6-$\sigma$ detection which was coincident with both a GLEAM source (J003408-072153) and a 4.5-$\sigma$ detection in the standard image.
This source, which happens to be a known pulsar, is not included in our catalogue but is discussed separately in Section~\ref{sec:pulsar}.

\subsection{Scintillation indices}
In order to understand the sub-arcsecond properties of our sources, scintillation indices are required (i.e. the scintillating flux density as a fraction of the mean flux density of the source). 

\subsubsection{Deriving Scintillating Flux Densities and upper limits} \label{Sec:ScintFlux}
As detailed in paper I, the values extracted from the variability image have to be transformed in order to recover the scintillating flux density $\Delta S_{scint}$:
\begin{equation}
	\Delta S_{scint} = \sqrt{P^2 -\mu^2} .
	\label{eqn:delta_s_s}
\end{equation}
We derive the error on $\Delta S_{scint}$ taking into account both the system noise, and the error due to the finite number of samples in the time series exactly as described in paper I.

For sources without detections in the variability image, we calculate an upper limit on the scintillating flux density from the measured value in the variability image at the position of the source and the 2-$\sigma$ error value:
\begin{equation}
	\max(\Delta S_{\rm scint}) = \sqrt{\left(P+2\sigma\right)^2 - \mu^2} .
	\label{eqn:delta_s_upper}
\end{equation}
We do not report these values directly, rather they are used to derive the scintillation indices (or upper limits thereon).

\subsubsection{Deriving the mean flux density}
In contrast to Paper I, all of the sources in our catalogue are detected in our standard image, and so we use this image to determine the mean flux density of each of our sources.
Using the standard image as the denominator of the scintillation index has the advantage that any instrumental errors, such as errors in the primary beam model, cancel exactly.
As shown in Paper I, it is possible that the centroid of the scintillating flux density is not coincident with the corresponding continuum source, due to arcminute-scale source structure.
To mitigate this, we obtained the corresponding flux density in the standard image of each component identified in the variability image by running Aegean in ``priorized'' mode (priorize=3).
Aegean then returns the peak fitted flux density within half power of the major and minor axis\footnote{https://github.com/PaulHancock/Aegean}.
The fractional error on the scintillation index for each source was then estimated in the standard way (using the fractional uncertainty on the scintillating flux density, added in quadrature with the fractional uncertainty in flux density from the standard image). 

\subsection{Solar elongation corrections}
\label{sec:elongation}
\begin{center}
\begin{figure}
\includegraphics[scale=0.34, angle=0]{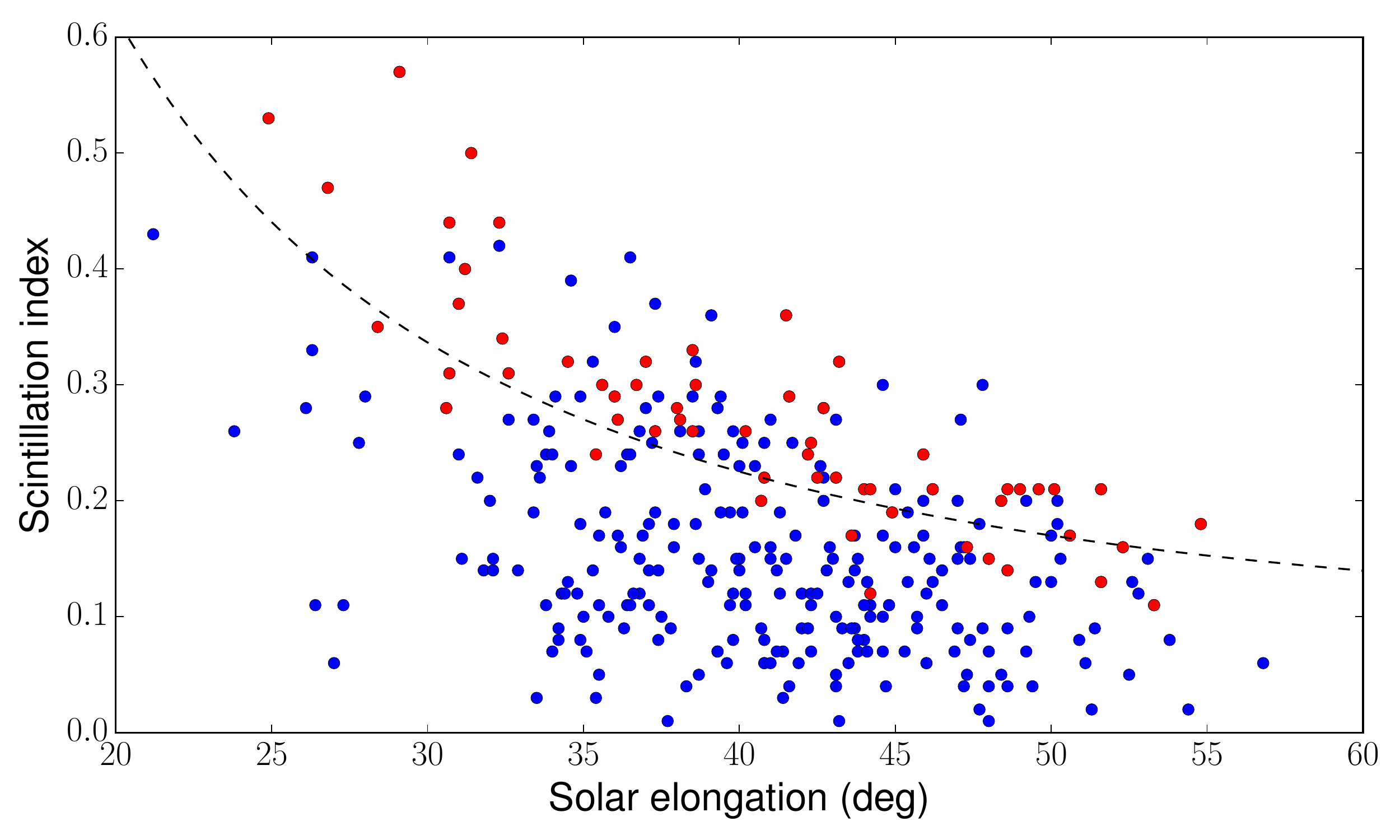}
\caption{Scintillation index as a function of solar elongation for all sources in our sample where IPS was detected. The dotted line shows the empirical fit to the average  profile of scintillation indices from \protect \cite{Rickett1973} for compact sources as they move across different solar elongations. The red dots show flat- or inverted-spectrum radio sources (spectral index $\geq$ -0.4), which are expected to be compact.}  
\label{Fig:scintil-Elong2}
\end{figure}
\end{center}

Figure \ref{Fig:scintil-Elong2} plots the scintillation index against solar elongation for all sources in our field with an IPS detection. The scintillation index is affected by source structure (more extended sources will have lower scintillation indices: see Section~\ref{sec:angularSize}). Changing solar wind conditions along each line of sight also vary the scintillation index stochastically. Our data also show that the scintillation index varies with solar elongation. In the weak scintillation regime, the average scintillation index of a point source ($m_{pt}$) approximately follow a simple empirical relationship \citep{Hewish1969,Rickett1973}.
\begin{equation}
	m_{pt}=0.06\lambda^{1.0}p^{-1.6}
	\label{eqn:rickett}
\end{equation}
where $\lambda$ is the wavelength of observation, and $p$ is the point of closest approach of the line of sight to the Sun in AU (the sine of the solar elongation of the source). This relation is shown as a dashed line in Figure \ref{Fig:scintil-Elong2}. The sources across our entire field are expected to be in the weak scintillation regime (i.e. $m_{pt}<1$). Note that the power law index is an average, and has been shown to change slightly depending on the source and on the solar wind conditions \citep{Manoharan1993SoPh..148..153M}.

We can test that our own data approximate this relationship by examining the behaviour of a set of sources likely to be compact.
Since most flat spectrum sources are expected to be compact AGN cores, it is instructive to see where these sources lie. Flat spectrum sources $\alpha\geq-0.4$ are indicated by the red dots in Figure \ref{Fig:scintil-Elong2}.   
These clearly lie among the sources with the highest scintillation index, and follow the expected trend. The up-scattering of some objects above the dotted line is due to solar wind variation.

\subsubsection{Normalised scintillation indices}
\label{sec:nsi}
To compare the scintillation properties of all the sources in our field in a meaningful way, we need to remove the effect of solar elongation. To do this, we divided the observed scintillation index by that predicted by Equation \ref{eqn:rickett}. We include this \textit{Normalised Scintillation Index (NSI)}\ in Table \ref{Tab:MainTable}. For sources where IPS was not detected, we normalised the upper limit by the same procedure. Sources dominated by a sub-arcsec compact component should have NSI $\sim$1.

\subsubsection{Normalised S/N}
\label{sec:nsn}
For statistical comparisons of scintillators and non-scintillators it is also convenient to define a sub-sample for which scintillation \emph{could have} been detected at a particular signal to noise ratio. Whether scintillation can be detected for a particular compact source depends on the sensitivity of our variability image at that point on the sky (which is strongly correlated with the sensitivity in the standard image) as well as the solar elongation of the source.

To take account of these biases, we define an adjusted S/N ratio that accounts for the dependence of scintillation index on distance from the Sun. 
This \textit{Normalised S/N} is calculated from the signal to noise ratio in the standard image by multiplying by the factor given by Equation~\ref{eqn:rickett}. 

\begin{center}
\begin{figure}
\includegraphics[scale=0.35, angle=0]{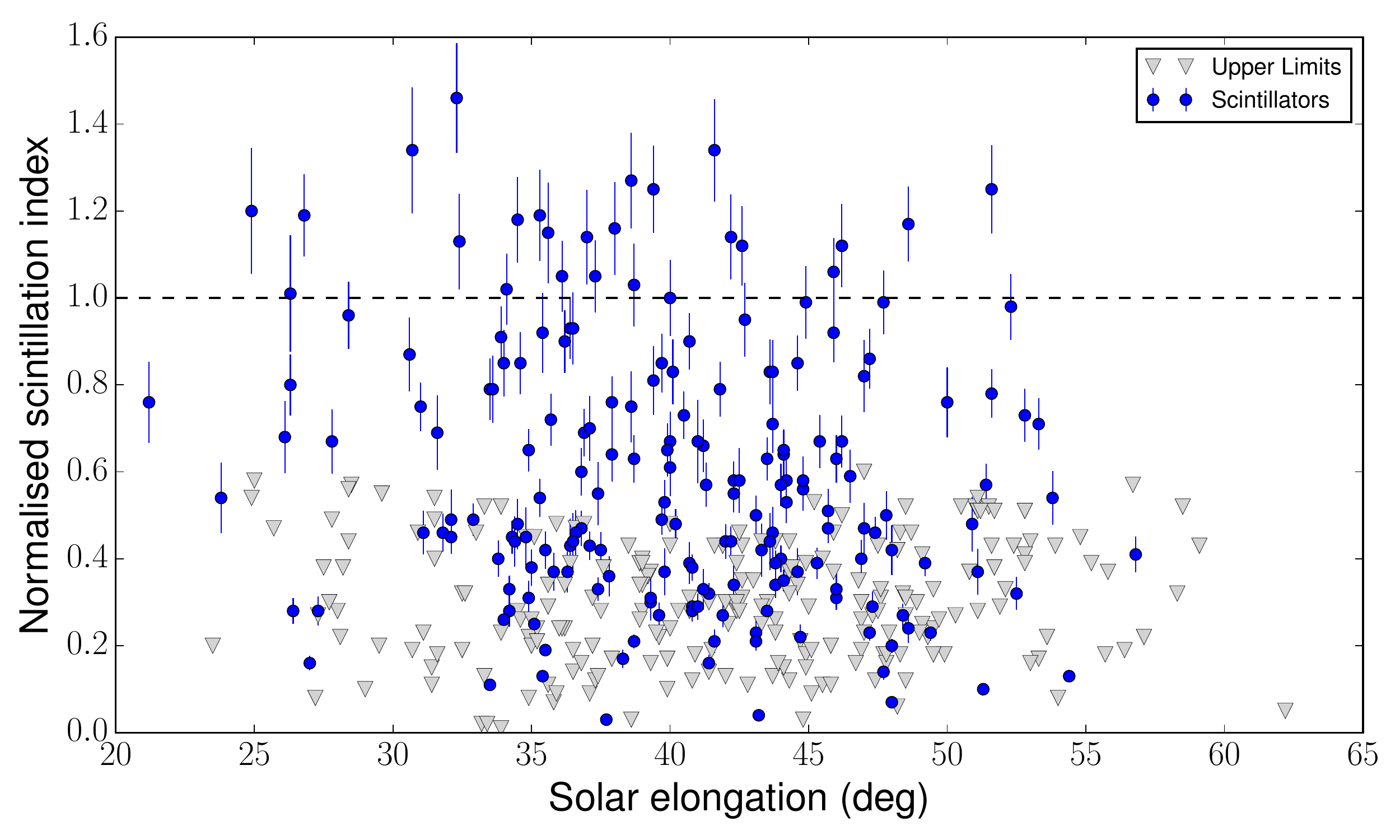}
\caption{Scintillation indices, normalised using the fit according to Eq. \ref{eqn:rickett} and as shown in Figure \ref{Fig:scintil-Elong2}, are plotted as a function of solar elongation for the high S/N subsample. Scintillating sources are shown by blue filled circles with error bars, and grey triangles show the upper limits for sources where scintillation was not detected. }
\label{Fig:scintil-Elong_NSI}
\end{figure}
\end{center}

For later analysis, we define a `high S/N' sub-sample to be those sources (scintillating or with limits) that have normalised S/N  $\geq$ 12.5 (see~\ref{sec:high_sn_subsample}).
Figure \ref{Fig:scintil-Elong_NSI} shows the normalised scintillation index (NSI) for all sources with normalised S/N  $\geq$ 12.5, plotted against solar elongation, showing that the IPS detection limit is now close to uniform over the whole observed field.

\subsection{Angular sizes of scintillating sources}
\label{sec:angularSize}
The critical scale for a source in the regime of weak scintillation (see Section~\ref{sec:elongation}) is the angle subtended by the Fresnel scale, given by 
\begin{eqnarray}
\theta_{\rm F} = \sqrt{\frac{\lambda}{2 \pi D} } = 0.30 \left( \frac{\nu}{150\,{\rm MHz}} \right)^{-1\mathbin{/}2} \left( \frac{D}{1\,{\rm AU}} \right)^{-1/2} {\rm arcsec}, \label{thetaFeq}
\end{eqnarray}
where $D$ is the effective distance to the scattering material.  A useful approximation to $D$ is the distance between the Earth and the point of closest approach of the line of sight to the source from the Sun \citep{Little1966}.  For a source containing structure whose size lies below the Fresnel angle, the corresponding characteristic timescale associated with the scintillations is
\begin{eqnarray}
t_{\rm F} = \frac{D \theta_{\rm F}}{v_{\rm scint}} 
= 0.55 \left( \frac{\nu}{150\,{\rm MHz}} \right)^{-1/2} \left( \frac{D}{1\,{\rm AU}} \right)^{1/2} \left( \frac{v_{\rm scint}}{400\,{\rm km\,s}^{-1}} \right) {\rm s},
\end{eqnarray}
where we have normalised to a fiducial scintillation speed of $400\,$km\,s$^{-1}$ that is characteristic of the solar wind emanating from the equatorial regions of the Sun.  Such a `point' source exhibits a scintillation index (i.e.\,\,fractional root-mean-square variations) with an amplitude
\begin{eqnarray}
m_{\rm pt} = \left( \frac{\theta_{\rm F}}{\theta_{\rm diff}} \right)^{5/6}, \label{mpteqn}
\end{eqnarray}
where $r_{\rm diff} = D \theta_{\rm diff}$ is the diffractive scale, the transverse scale over which inhomogeneities in the interplanetary plasma impart one radian of root-mean-square phase difference in the wavefront of the scintillating source.  The index of $5/6$ in Eq.(\ref{mpteqn}) follows as a result of the assumption that the turbulence follows a Kolomogorov spectrum \citep[see, e.g.,][]{Narayan1992}.  Since $r_{\rm diff}$ is not known {\it a priori}, it must either be measured or estimated; the dotted line in Figure \ref{Fig:scintil-Elong2} represents the empirical estimate of $m_{\rm pt}$ used throughout this paper.

A source whose compact structure substantially exceeds $\theta_{\rm F}$ 
exhibits a characteristic timescale set by the source size, $\theta_{\rm src}$, that exceeds the timescale, $t_{\rm F}$:
\begin{eqnarray}
t_{\rm scint} \approx \frac{D \theta_{\rm src}}{v_{\rm scint}} = 1.8 \left(\frac{\theta_{\rm src}}{1\,{\rm arcsec}} \right) 
\left( \frac{D}{1\,{\rm AU}} \right) 
\left( \frac{v_{\rm scint}}{400\,{\rm km\,s}^{-1}}  \right)\,{\rm s}.
\end{eqnarray}
The scintillation index of an extended source, $m$, is reduced by a ratio $\approx (\theta_{\rm src}/\theta_{\rm F})$ relative to a point source, so that the NSI is a direct measure of the source angular size relative to the Fresnel angle:
\begin{eqnarray}
{\rm NSI} = \frac{m}{m_{\rm pt}} = \left( \frac{\theta_{\rm F}}{\theta_{\rm src}} \right), \qquad \theta_{\rm src} > \theta_{\rm F}. \label{NSIeq}
\end{eqnarray}
We stress that this relationship between source size and the NSI applies to a source consisting of a single component; the NSI of a source that possesses an additional, more extended component would be further reduced.  

For sources closer to the Sun the scintillations enter the regime of strong scattering, for which refractive scintillation is the dominant source of intensity fluctuations.  Formally, this regime applies when the diffractive scale $r_{\rm diff}$ is substantially smaller than the Fresnel scale, $r_{\rm F} = D \theta_{\rm F}$.  Refractive scintillation is not relevant for the solar elongations probed in the present dataset, and is not discussed further here.  

Daily changes in the solar weather can produce small changes in the parameters $r_{\rm diff}$ and the effective distance to the scattering region, and hence, $r_{\rm F}$, with commensurate changes to the modulation index (see \citet{Kaplanetal2015} for an extreme manifestation of this).  

\begin{center}
\begin{figure}
\includegraphics[scale=0.42, angle=0]{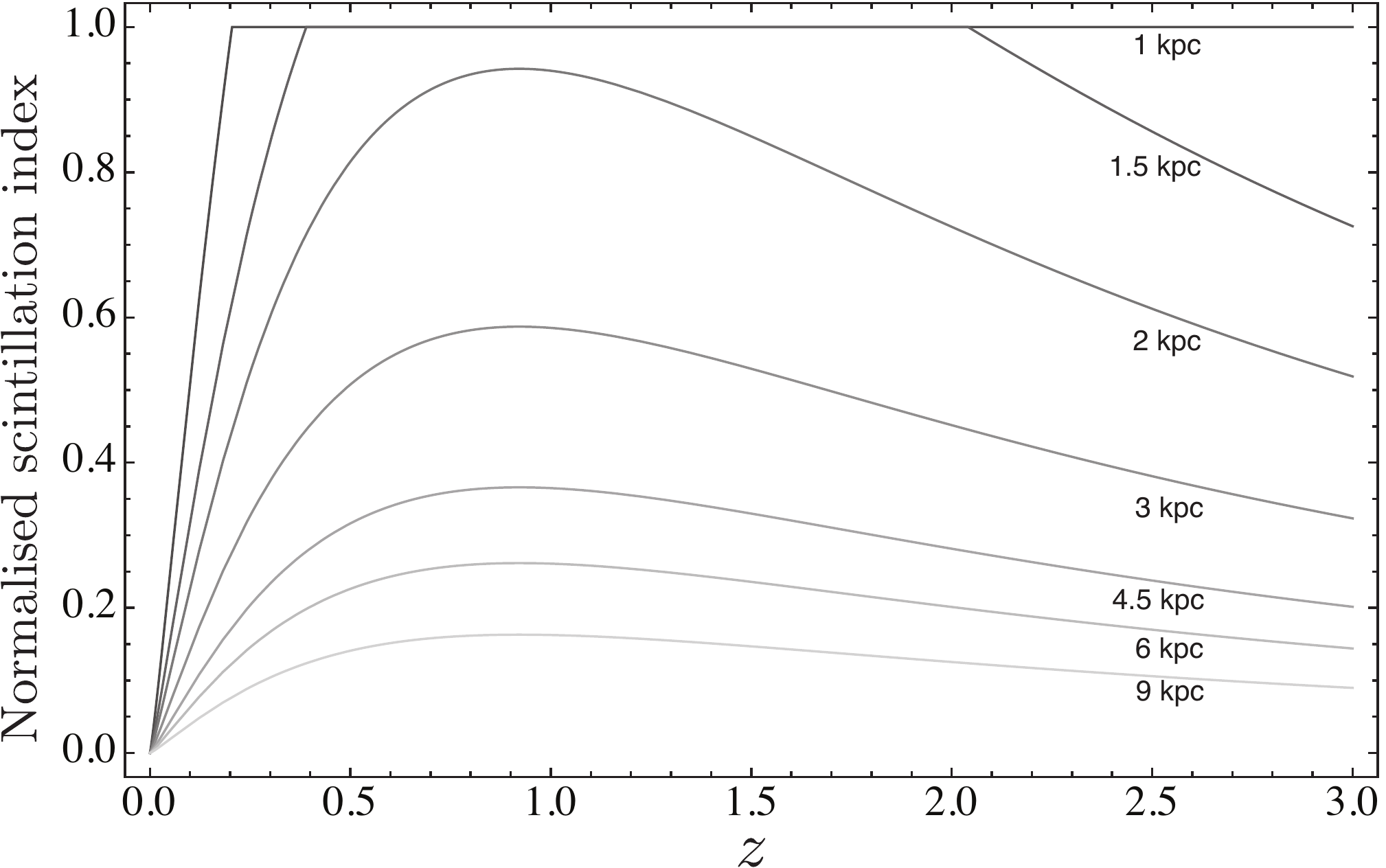}
\caption{The predicted normalised scintillation index (Y-axis) based on Eq. \ref{NSIeq} for objects with different linear sizes, as a function of redshift for scintillation at 150\,MHz and at an effective scattering distance $D=1\,$AU (i.e.\,\,the fiducial values in Eq. \ref{thetaFeq}). For AGN with linear size $\lesssim$0.5 kpc \protect{\citep[figure 2 in][]{Chhetri2012}}, we expect the normalised scintillation index to be close to 1 for redshifts above 0.2. }
\label{Fig:scint-Z}
\end{figure}
\end{center}

Extragalactic sources with components smaller than a few kiloparsec (cores and hotspots) are expected to scintillate at all redshifts greater than about 0.3, as shown in Figure \ref{Fig:scint-Z}.

\subsection{Catalogue of sources with scintillation index measurements}
Table \ref{Tab:MainTable} lists 25 sources from the central region of our field, in ascending order of their RA. The full table will be available as supplementary material to this publication, in on-line form through \textit{VizieR} catalogue access tool \citep{Ochsenbein2000}, and also upon request to the authors.
The columns provided in Table \ref{Tab:MainTable} are as follows:

\begin{table*}
\footnotesize
\centering
{
\setlength\tabcolsep{2.0pt} 
\resizebox{\textwidth}{!}{%
\begin{tabular}
{|c|c|c|c|c|c|l|c|c|r|c|c|c|c|r|c|r|r|}
  \hline
  \multicolumn{1}{|c|}{GLEAM Name} &
  \multicolumn{1}{c|}{RA} &
  \multicolumn{1}{c|}{Dec} &
  \multicolumn{1}{c|}{S$_{162}$} &
  \multicolumn{1}{c|}{$\pm$} &
  \multicolumn{1}{c|}{$\alpha$} &
  \multicolumn{1}{c|}{$\alpha$} &
  \multicolumn{1}{c|}{$\pm$} &
  \multicolumn{1}{c|}{N} &
  \multicolumn{1}{c|}{S$_{1400}$} &
  \multicolumn{1}{c|}{Solar Elong.} &
  \multicolumn{1}{c|}{Norm.} &
  \multicolumn{1}{c|}{Scint.} &
  \multicolumn{1}{c|}{$\pm$} &
  \multicolumn{1}{c|}{Norm.} &
  \multicolumn{1}{c|}{$\pm$} &
  \multicolumn{1}{c|}{Flags}\\
  
  \multicolumn{1}{|c|}{} &
  \multicolumn{1}{|c|}{J2000} &
  \multicolumn{1}{|c|}{J2000} &
  \multicolumn{2}{l|}{(Jy/beam)} &
  \multicolumn{1}{c|}{GLEAM} &
  \multicolumn{1}{c|}{Flag} &
  \multicolumn{1}{c|}{} &
  \multicolumn{1}{c|}{NVSS} &
  \multicolumn{1}{c|}{(Jy)} &
  \multicolumn{1}{c|}{(deg)} &
  \multicolumn{1}{c|}{S/N} &
  \multicolumn{1}{c|}{index} &
  \multicolumn{1}{c|}{} &
  \multicolumn{1}{c|}{scint. index} &
  \multicolumn{1}{c|}{} &
  \multicolumn{1}{c|}{}\\
  
  \multicolumn{1}{|c|}{(1)} &
  \multicolumn{1}{c|}{(2)} &
  \multicolumn{1}{c|}{(3)} &
  \multicolumn{1}{c|}{(4)} &
  \multicolumn{1}{c|}{(5)} &
  \multicolumn{1}{c|}{(6)} &
  \multicolumn{1}{c|}{(7)} &
  \multicolumn{1}{c|}{(8)} &
  \multicolumn{1}{c|}{(9)} &
  \multicolumn{1}{c|}{(10)} &
  \multicolumn{1}{c|}{(11)} &
  \multicolumn{1}{c|}{(12)} &
  \multicolumn{1}{c|}{(13)} &
  \multicolumn{1}{c|}{(14)} &
  \multicolumn{1}{c|}{(15)} &
  \multicolumn{1}{c|}{(16)} &
  \multicolumn{1}{c|}{(17)} \\
\hline
  J003822-224320 & 00:38:22.36 & -22:43:20.88 & 0.760 & 0.016 &-0.94 & ... & 0.14 & 1 & 0.086 & 47.2 &  8.31 & $<$0.09 &  ...  & $<$0.51 &  ...  & ... \\
  J003824-225256 & 00:38:24.86 & -22:52:56.09 & 2.327 & 0.016 &-0.74 & ... & 0.08 & 1 & 0.391 & 47.4 & 29.74 &  0.08 & 0.007 &  0.46 & 0.037 & ... \\
  J003829-211957 & 00:38:29.77 & -21:19:57.41 & 1.081 & 0.017 & 0.15 & ... & 0.15 & 1 & 0.828 & 46.2 & 13.91 &  0.21 & 0.018 &  1.12 & 0.096 &  AP \\
  J003830-202251 & 00:38:30.90 & -20:22:51.79 & 0.995 & 0.017 &-0.85 & ... & 0.13 & 1 & 0.139 & 45.4 & 12.44 & $<$0.08 &  ...  & $<$0.40 &  ...  & ... \\
  J003837-171556 & 00:38:37.67 & -17:15:56.64 & 0.843 & 0.016 & -0.91 & ... & 0.12 & 1 & 0.100 & 43.1 & 12.13 & $<$0.11 &  ...  & $<$0.54 &  ...  & ... \\
  J003849-222521 & 00:38:49.83 & -22:25:21.61 & 0.879 & 0.016 & -0.84 & ... & 0.12 & 1 & 0.133 & 46.9 & 12.39 & $<$0.09 &  ...  & $<$0.47 &  ...  & ... \\
  J003851-202835 & 00:38:51.87 & -20:28:35.27 & 0.215 & 0.017 & -0.81 & ... & 0.55 & 1 & 0.030 & 45.4 &  2.98 & $<$0.16 &  ...  & $<$0.82 &  ...  & ... \\
  J003906-242506 & 00:39:06.77 & -24:25:06.66 & 1.289 & 0.016 & -0.99 & ... & 0.09 & 1 & 0.217 & 48.5 & 15.86 & $<$0.05 &  ...  & $<$0.27 &  ...  & ... \\
  J003907-184629 & 00:39:07.87 & -18:46:29.49 & 0.311 & 0.015 & -1.02 & ... & 0.25 & 2 &  ...  & 44.1 &  4.76 & $<$0.17 &  ...  & $<$0.86 &  ...  & ... \\
  J003908-221953 & 00:39:08.35 & -22:19:53.80 & 0.285 & 0.015 & -0.63 & ... & 0.26 & 1 & 0.114 & 46.8 &  3.87 & $<$0.13 &  ...  & $<$0.73 &  ...  &  AF \\
  J003911-203735 & 00:39:11.84 & -20:37:35.05 & 0.497 & 0.017 & -0.68 & ... & 0.24 & 1 & 0.112 & 45.5 &  5.66 & $<$0.18 &  ...  & $<$0.93 &  ...  & ... \\
  J003919-190616 & 00:39:19.42 & -19:06:16.70 & 0.295 & 0.015 & -0.89 & ... & 0.32 & 1 & 0.054 & 44.3 &  3.37 & $<$0.10 &  ...  & $<$0.50 &  ...  & ... \\
  J003922-203521 & 00:39:22.78 & -20:35:21.67 & 0.413 & 0.017 & -0.87 & ... & 0.25 & 1 & 0.078 & 45.5 &  4.92 & $<$0.14 &  ...  & $<$0.74 &  ...  & ... \\
  J003925-170300 & 00:39:25.79 & -17:03:00.70 & 0.457 & 0.015 & -0.70 & ... & 0.20 & 1 & 0.092 & 42.8 &  6.67 & $<$0.16 &  ...  & $<$0.77 &  ...  & ... \\
  J003941-225703 & 00:39:41.99 & -22:57:03.85 & 0.576 & 0.016 & -0.67 & ... & 0.18 & 1 & 0.130 & 47.2 &  7.00 & $<$0.16 &  ...  & $<$0.89 &  ...  & ... \\
  J003945-155153 & 00:39:45.75 & -15:51:53.16 & 0.442 & 0.017 & -0.95 & ... & 0.20 & 1 & 0.051 & 41.9 &  5.40 & $<$0.12 &  ...  & $<$0.57 &  ...  & ... \\
  J004000-172206 & 00:40:00.39 & -17:22:06.66 & 0.160 & 0.015 & -0.72 & ... & 0.59 & 1 & 0.035 & 42.9 &  3.57 & $<$0.17 &  ...  & $<$0.82 &  ...  & ... \\
  J004008-230142 & 00:40:08.78 & -23:01:42.78 & 0.379 & 0.016 & -0.86 & ... & 0.23 & 1 & 0.056 & 47.2 &  4.26 & $<$0.04 &  ...  & $<$0.21 &  ...  & ... \\
  J004011-174942 & 00:40:11.62 & -17:49:42.88 & 0.497 & 0.015 & -0.88 & ... & 0.19 & 1 & 0.087 & 43.3 &  6.70 & $<$0.05 &  ...  & $<$0.25 &  ...  & ... \\
  J004012-201114 & 00:40:12.02 & -20:11:14.11 & 0.291 & 0.017 & -0.48 & ... & 0.41 & 1 & 0.110 & 45.0 &  3.71 & $<$0.13 &  ...  & $<$0.65 &  ...  & ... \\
  J004018-213141 & 00:40:18.84 & -21:31:41.18 & 0.656 & 0.016 & -0.59 &  *  & ...  & 1 & 0.149 & 46.0 &  9.03 & $<$0.10 &  ...  & $<$0.54 &  ...  & ... \\
  J004026-210924 & 00:40:26.99 & -21:09:24.66 & 0.525 & 0.017 & -0.86 & ... & 0.2  & 1 & 0.081 & 45.7 &  7.17 & $<$0.10 &  ...  & $<$0.52 &  ...  & ... \\
  J004029-204651 & 00:40:29.95 & -20:46:51.41 & 0.243 & 0.017 & -0.63 & ... & 0.41 & 1 & 0.040 & 45.4 &  3.10 & $<$0.20 &  ...  & $<$1.03 &  ...  & ... \\
  J004043-191830 & 00:40:43.83 & -19:18:30.97 & 0.414 & 0.016 & -1.06 & ... & 0.23 & 1 & 0.042 & 44.3 &  4.67 & $<$0.14 &  ...  & $<$0.69 &  ...  & ... \\
  J004048-204329 & 00:40:48.06 & -20:43:29.42 & 1.710 & 0.017 & -0.53 & ... & 0.09 & 1 & 0.641 & 45.3 & 21.22 & 0.07 & 0.007 &  0.39 & 0.036 &   A \\
  J004049-175706 & 00:40:49.65 & -17:57:06.36 & 0.290 & 0.015 & -0.75 & ... & 0.31 & 1 & 0.063 & 43.2 &  4.34 & $<$0.13 &  ...  & $<$0.62 &  ...  & ... \\

\hline
\end{tabular}
}
}
\caption{As an example of the data format, this table shows 25 sources from central region in our MWA field. Flag '*' in Column 7 means the spectra index provided in Column 6 is an estimate and not from GLEAM catalogue. "$<$" in columns 13 and 15 indicate upper limits for non-scintillating sources. Flags 'A', 'F', and 'P' in Column 17 mean that the source has an AT20G counterpart, is a \fermi\ source, and is a peaked-spectrum source respectively.}
\label{Tab:MainTable}
\end{table*}

\begin{tabbing}
	$(1)$ ~~~\= GLEAM name for the source.\\
    $(2)$	\> Right Ascension (GLEAM).\\
    $(3)$	\> Declination (GLEAM).\\
	$(4)$   \> Flux density at 162\,MHz, interpolated from the GLEAM\\ \> 158 and 166\,MHz peak flux densities(Jy/beam).\\
	$(5)$   \> Error in 162\,MHz flux density estimated using local\\\> RMS.\\
	$(6)$   \> Radio spectral index across the GLEAM band \\\>(72-231\,MHz). \\
    $(7)$   \> Flag for spectral index in GLEAM band. '*' indicates \\\>that GLEAM does not provide a spectral index for \\\>these sources which are not consistent with a pure \\\>power law.  We have provided an approximate power-\\\>law fit to the GLEAM data.\\
	$(8)$   \> Error in spectral index.\\
    $(9)$	\> Number of NVSS sources within 60\,arcsec of\\ \>the GLEAM source position.\\
	$(10)$   \> Integrated flux density from NVSS (Jy). No NVSS flux\\ \> density is listed if more than one counterpart (Col. 6)\\\>is found within search radius.\\
    $(11)$   \> Solar elongation for the source (degrees).\\
    $(12)$   \> Normalised S/N ratio based on standard image S/N.\\
	$(13)$  \> Scintillation index, or an upper limit for\\ \>non-scintillating sources. \\
	$(14)$  \> Error in scintillation index. \\
	$(15)$  \> Normalised scintillation index, or an upper limit \\ \>for non-scintillating sources.\\
	$(16$  \> Error in normalised scintillation index. \\
	$(17)$  \> Flag: \\
    		  \>\textit{\rm A : AT20G source \citep{Murphy2010}}\\
		  \>\textit{\rm F : \fermi\ gamma-ray source \citep[3FGL;][]{Acero2015}}\\
		  \>\textit{\rm P : MWA peaked spectrum source \citep{Callinghametal17} } \\
\end{tabbing}

For each object in the catalogue, we also present SED plots covering the frequency range between 72 MHz (GLEAM) and 20 GHz (AT20G), where available, as supplementary material available to this publication, in on-line form. 

\section{Discussion and Analysis}
\label{sec:discussion}

\subsection{IPS classes and the high S/N subsample}\label{Sec:characterisingIPS}
\label{sec:high_sn_subsample}

\noindent
For the purpose of analysis, we now split our sample into three broad classes based on their normalised scintillation index (NSI): 

\begin{itemize}
\item 
{\bf Strongly scintillating}\ sources with NSI $\geq$ 0.9 (111 objects in Table 1). These are objects where the low-frequency flux density is dominated by a single sub-arcsec component.   
\item 
{\bf Moderately scintillating} sources with 0.4 $\leq$ NSI $<$ 0.9 (132 objects in Table 1). Members of this class may include compact double sources, as well as sources where a single sub-arcsec component is embedded within more extended low-frequency emission.
\item 
Sources with {\bf weak or no scintillation} with NSI $<$ 0.4 (389 objects in Table 1. These 389 objects include 59 objects with IPS detections and 330 objects with NSI upper limits less than to 0.40, so they can be unambiguously classified as weak or non-scintillating sources). In these sources, the observed low-frequency radio emission is expected to be dominated by extended rather than compact components. 
\end{itemize}

A total of 632 sources (25\% of the full sample listed in Table 1) can be classified unambiguously in this way. A further 811 sources have upper limits on NSI that are between 0.4 and 0.9, so these objects could plausibly have either moderate or weak/no scintillation. Most of these sources (93\%) have normalised S/N $<$ 12.5.  The remaining 1107 sources 
have upper limits on NSI that are larger than 0.9, so no meaningful statement about their scintillation properties can be made from our current dataset but all these objects also have normalised S/N $<$ 12.5 so can be excluded from a high S/N sub sample. 

We chose to define a `high S/N' subsample using a uniform S/N cut at normalised S/N $\geq$ 12.5.  
This cut-off was chosen to allow a reasonably unambiguous classification of the scintillation properties of 
individual sources, while still providing a large enough sample (414 objects) for  statistical analysis. 
The median 162\,MHz flux density is 2.07\,Jy for the 414 objects with normalised S/N $\geq$ 12.5, and 0.54\,Jy for the remaining sources in Table 1 with normalised S/N $<$ 12.5. 
For the rest of this paper, we restrict our analysis to the high S/N subsample of 414 sources with normalised S/N $\geq$ 12.5 unless otherwise stated.  

\begin{center}
\begin{figure}
\includegraphics[scale=0.33, angle=0]{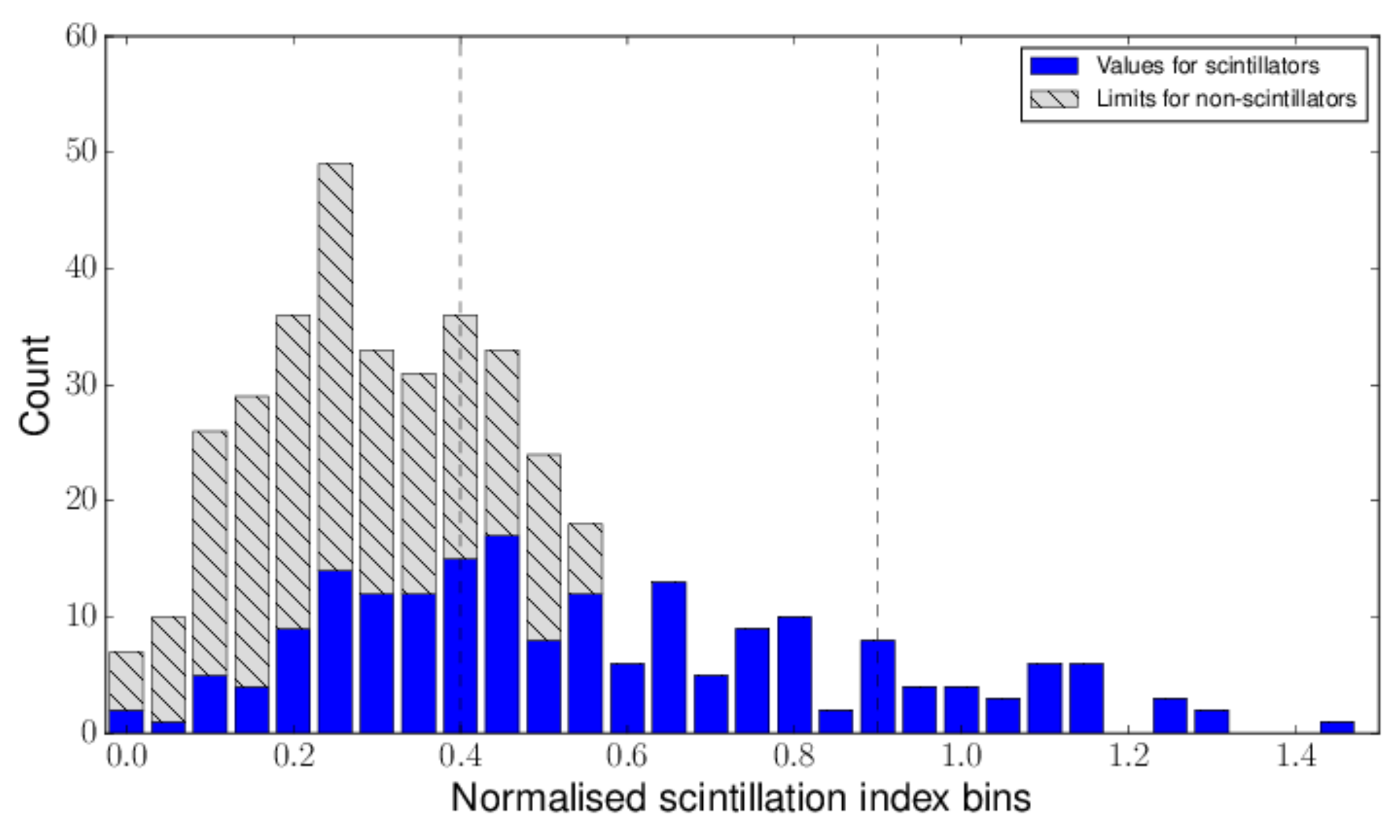}
\caption{Histogram of normalised scintillation indices of sources showing scintillation in our observation field, and upper limits for non-scintillating sources (high S/N sample). Vertical lines are drawn at 0.9 and 0.4, where we make distinction in different populations.} 
\label{histo:scintillationIndices}
\end{figure}
\end{center}

Figure \ref{Fig:scintil-Elong_NSI} plots the distribution of normalised scintillation indices (and limits) as a function of solar elongation for the high S/N subsample. The NSI upper limits for undetected sources in this sub-sample are now roughly uniform with solar elongation. 
Figure \ref{histo:scintillationIndices} shows a histogram of normalised scintillation indices (and upper limits) for the high S/N sources with the boundaries of our three scintillation classes marked. All NSI upper limits are now below 0.6, allowing a reasonably clear separation into scintillation classes as summarised in Table \ref{Tab:highsN_all}. 

The overall IPS detection rate for the high S/N subsample is 47\%, and 9\% of the high S/N sources are strong scintillators. 
The scintillation properties are summarised in Table \ref{Tab:highsN_all}, which allows us to conclude that at least 32\% of bright low-frequency radio sources have a sub-arcsec compact component that contributes 40\% or more of the total flux density at 162\,MHz.

\begin{table}
\resizebox{\columnwidth}{!}{%
\begin{tabular}{llrr}
\hline            
\multicolumn{1}{l}{Class} & \multicolumn{1}{l}{Definition } & \multicolumn{1}{c}{Sources} & Fraction \\
\hline
Strong scintillators   & NSI$\geq$0.90   &	37   &  8.9\% \\
Moderate scintillators & 0.40$\leq$NSI$<$0.90   & 97 & 23.4\%  \\
Weak/non-scintillators  & NSI$<$0.40   & 221 & 53.4\% \\
\multicolumn{2}{l}{NSI upper limit unrestrictive (range 0.4--0.6) } & 59 & 14.3\% \\
&&& \\
Total & & 414 & \\
\hline
\end{tabular}
}
\caption{Summary of the overall scintillation properties of the 414 sources in our high S/N subsample (normalised S/N $\geq$ 12.5). }
\label{Tab:highsN_all}
\end{table}

\subsection{Source structure in the 1.4\,GHz VLA FIRST survey}
Part of our IPS field overlaps with the area covered by the 1.4\,GHz VLA FIRST survey \citep{Becker1995}. This allows us to compare the arcsec-scale structure imaged at 1.4\,GHz with that inferred from our IPS observations at 162\,MHz. Although the two observing frequencies are very different, this provides a useful first-order consistency check for our IPS results.  
As can be seen from Figure \ref{Fig:SurveysAngRes}, FIRST (with a 5\,arcsec beam) has the highest angular resolution of any of the large-area imaging surveys. The FIRST catalogue lists deconvolved major and minor axes (Maj, Min) for each source, and the accompanying catalogue description notes that sources with Maj $<$ 2\,arcsec can generally be regarded as unresolved. 

We examined the FIRST catalogue and images for the 89 MWA  sources in the high S/N subsample that lay within the FIRST overlap area. Of these 89 sources, 37 have a single FIRST component, 32 are resolved doubles in FIRST and 19 are associated with three or more FIRST components. One source, the steep-spectrum cluster relic GLEAM J004130-092221 (MRC\,0038-096), was resolved out and undetected in the FIRST survey. 

\begin{table}
\resizebox{\columnwidth}{!}{%
\begin{tabular}{llrrrrrc}
\hline            
 & & \multicolumn{4}{c}{FIRST components } \\
\multicolumn{1}{l}{Class} & \multicolumn{1}{l}{Definition } & \multicolumn{1}{c}{1} & \multicolumn{1}{c}{2} & \multicolumn{1}{c}{3+} & \multicolumn{1}{c}{0} & \multicolumn{1}{c}{Total} \\     
\hline
Strong scintillators   & NSI$\geq$0.90   &	 14   & .. & .. & .. & 14 \\
Moderate scintillators & 0.40$\leq$NSI$<$0.90   & 13  & 4 & .. & .. & 17 \\
Weak/non-scintillators  & NSI$<$0.40   & 6  & 22 & 13 & 1 & 42 \\
\multicolumn{2}{l}{NSI limit unrestrictive} & 4 & 6 & 6 & .. & 16 \\
&&&&& \\
Total & & 37 & 32 & 19 & 1 & 89 \\
\hline
\end{tabular}
}
\caption{1.4\,GHz morphologies of the high S/N sources that lie within the area covered by the VLA FIRST survey. Sixteen sources with NSI upper limits above 0.4 were not assigned to a scintillation class and are listed separately in the table.}
\label{Tab:highsN_FIRST}
\end{table}

73 of the 89 high S/N sources in the FIRST overlap area can be split into scintillation classes as discussed in Section~\ref{sec:high_sn_subsample}. As can be seen from Table \ref{Tab:highsN_FIRST}, there is generally good agreement between our 162\,MHz IPS observations and the structures seen by FIRST at 1.4\,GHz. 
In particular, we note that the six weak/non-scintillating sources matched with a single FIRST component are all spatially-extended in the FIRST images, with deconvolved major axis sizes ranging from 6.0 to 8.5\,arcsec. 

The median normalised scintillation index at 162\,MHz is 0.79 for objects matched with single-component FIRST sources, and $<0.20$ for sources matched with two or more FIRST components. This again indicates a reassuring degree of consistency between the two indicators of small-scale source structure. 

\subsection{Radio SEDs for the three scintillation classes} \label{Sec:AvgSED}
The IPS properties of our sources can be used to investigate how the dominance of sub-arcsec compact components at 162\,MHz affects their radio SEDs. At higher frequencies, there is a well-known relationship between source size and spectral index, with flat-spectrum radio sources being significantly more compact than those with steeper radio spectra.
We first discuss the radio spectral index across the lower frequency GLEAM band (76-227\,MHz), then we consider the higher frequency radio spectral index between GLEAM observations at 162\,MHz and NVSS at 1.4\,GHz.   
The field we observed lies close to the south Galactic pole, so the source population observed is expected to be almost entirely extragalactic. 

\subsubsection{Radio SEDs across the GLEAM Band} 
\cite{Hurley-Walker2017} discuss the overall spectral properties of radio sources in the GLEAM catalogue, and measure a median spectral index of -0.83$\pm$0.11 across the GLEAM band (from 76--227\,MHZ) for sources brighter than 1.0\,Jy. This value is in good agreement with the median values of -0.82 measured by \cite{Lane2014} at 74-1400\,MHz and -0.83 measured by \cite{Mauch2003} at 843-1400\,MHz.   

For sources that are consistent with a power law the GLEAM catalogue provides a spectral index across the GLEAM band. These are provided in Column 6 in Table \ref{Tab:MainTable}. Sources for which no spectral index values are provided in the GLEAM catalogue, we made a least squares linear fit to the spectra and used this to estimate the spectral indices. These sources have been given a flag "*" in Column 7 in Table \ref{Tab:MainTable} to indicate that these are approximate estimates only.

\begin{center}
\begin{figure}
\includegraphics[scale=0.72, angle=0]{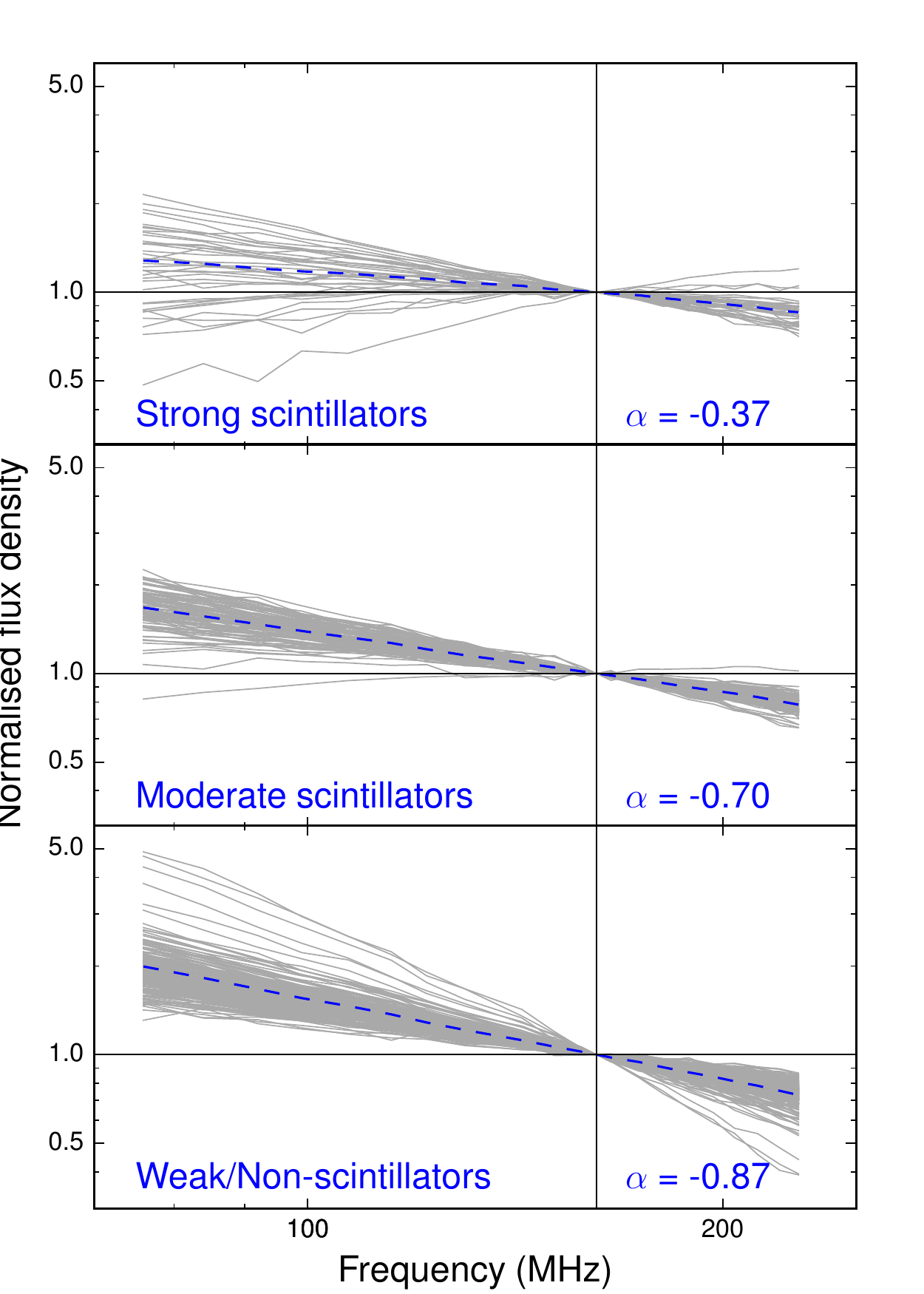}
\caption{Normalised SED of scintillating sources in the high S/N subsample for three class of objects. SEDs are drawn using flux densities from the GLEAM catalogue, normalised with their estimated peak flux density at 162 MHz (shown with a solid vertical line). The blue dashed lines represent the median SED for each class of objects. Two point spectral indices calculated for median SEDs of each class between the two frequency ends of the GLEAM survey are also shown.}
\label{Fig:avg-SEDmultiplot}
\end{figure}
\end{center}

We used the GLEAM catalogue to calculate a radio spectrum across the GLEAM band for each object in the high S/N subsample.  
Figure \ref{Fig:avg-SEDmultiplot} shows the GLEAM SEDs for individual objects in the three populations identified in Section  \ref{Sec:characterisingIPS}, along with the median SED for each class. For comparison each spectrum has been normalised to the 162\,MHz flux density listed in Table 1. The median spectral indices are also included in Table \ref{Tab:avgSED_stats}. 

\begin{table}
\resizebox{\columnwidth}{!}{%
\begin{tabular}{llrrrrrc}
\hline            
& & \multicolumn{4}{r}{Median radio spectral index} \\
\multicolumn{1}{l}{Class} & \multicolumn{1}{l}{Definition} &  \multicolumn{1}{c}{$\alpha_{76}^{277}$} & \multicolumn{1}{l}{N} & \multicolumn{1}{c}{$\alpha_{162}^{1400}$}& \multicolumn{1}{c}{N}  \\     
\hline
Strong scintillators   & NSI$\geq$0.90   &	 -0.37   & 37 & -0.68 & 37 \\
\multicolumn{2}{l}{\hspace{0.2cm}(Peaked sources only)} & -0.02 & 17 & -0.63 & 17 \\
\multicolumn{2}{l}{\hspace{0.2cm}(Non-peaked sources)}  & -0.60 & 20 & -0.83 & 20  \\
Moderate scintillators & 0.40$\leq$NSI$<$0.90   & -0.70  & 97 & -0.77 & 91\\
Weak/non-scintillators  & NSI$<$0.40    & -0.87   & 221 & -0.81 & 171 \\
\multicolumn{2}{l}{NSI limit unrestrictive} &  & 59 &  & 49 & \\
&&&& \\
Total   & & -0.82 & 414  & -0.80 & 347 \\ 
\hline
\end{tabular}
}
\caption{Median spectral indices for strong, moderate, and weak/non-scintillators for the high S/N subsample. Objects with unrestrictive NSI limits (above 0.40) are listed separately and were not included in the analysis. The median spectral index between 162 MHz and NVSS was only calculated for objects with a single NVSS counterpart, as reflected by the numbers in the rightmost column. }
\label{Tab:avgSED_stats}
\end{table}

As can be seen from Figure \ref{Fig:avg-SEDmultiplot}, the class of strongly-scintillating sources (NSI $\geq$ 0.90) includes a range of steep, flat, and rising/peaked spectra sources. For this population of compact sources, no single spectral class clearly dominates. The median spectral index across the GLEAM band (-0.37) is considerably flatter than the median value of -0.83 for all sources in the GLEAM catalogue. 

Among the extragalactic compact source population, flat-spectrum sources are expected to be cores of radio galaxies, QSOs or blazars. Compact steep spectrum sources, along with rising or peaked spectrum sources are thought to be either early stages in the evolution of a radio galaxies, or else frustrated from further development by their host environment \citep{Callingham2015}.

The moderately scintillating sources (0.40 $\leq$ NSI $<$ 0.9) in Figure \ref{Fig:avg-SEDmultiplot} are dominated by steep-spectrum radio sources. 
The median spectral index of this population (-0.70), is slightly flatter than the median spectral index of total sources in GLEAM. This can be understood as flattening of the overall spectrum of a radio galaxy by embedded flat- or rising spectra core, or hotspots with flat- or rising spectra. In the case of cores, these are starting to dominate the overall spectra and will become dominant at higher frequencies. In the case of hot-spots, their spectra becomes steep at higher frequencies, allowing us to distinguish between the hot-spots and cores. Alternatively, a class of population that can also show lowered normalised scintillation index and flat or rising spectra could also arise due to a new episode of activity in the core of a radio galaxy. Thus, this population provides a parent population that can be used to study regenerated activity of AGNs. 

Objects showing peaked spectra, with their peak below GLEAM frequencies will appear as compact steep spectrum (CSS) sources from higher frequency nomenclature. Such CSS objects embedded in larger structures, or hotspots in radio lobes can also be part of this population but their presence would not alter its average spectral index. Hence, the sources with lower normalised scintillation index are a composite of CSS sources, hot spots in radio lobes, and flat or rising spectra cores that are re-started or starting to dominate the overall spectra of the radio galaxy.  These are discussed further in Section \ref{Sec:PartiallyScintillatingSources}. 

The low or non-scintillating population is $\sim$ 53\% of our high S/N subsample and is dominated by steep-spectrum objects, and the median spectral index of this population (-0.87) is slightly steeper than the median for all GLEAM sources brighter than 1\,Jy. This is the expected behaviour of lobe dominated sources when no core or hotspots are present. 

\begin{center}
\begin{figure*}
\includegraphics[scale=0.50, angle=0]{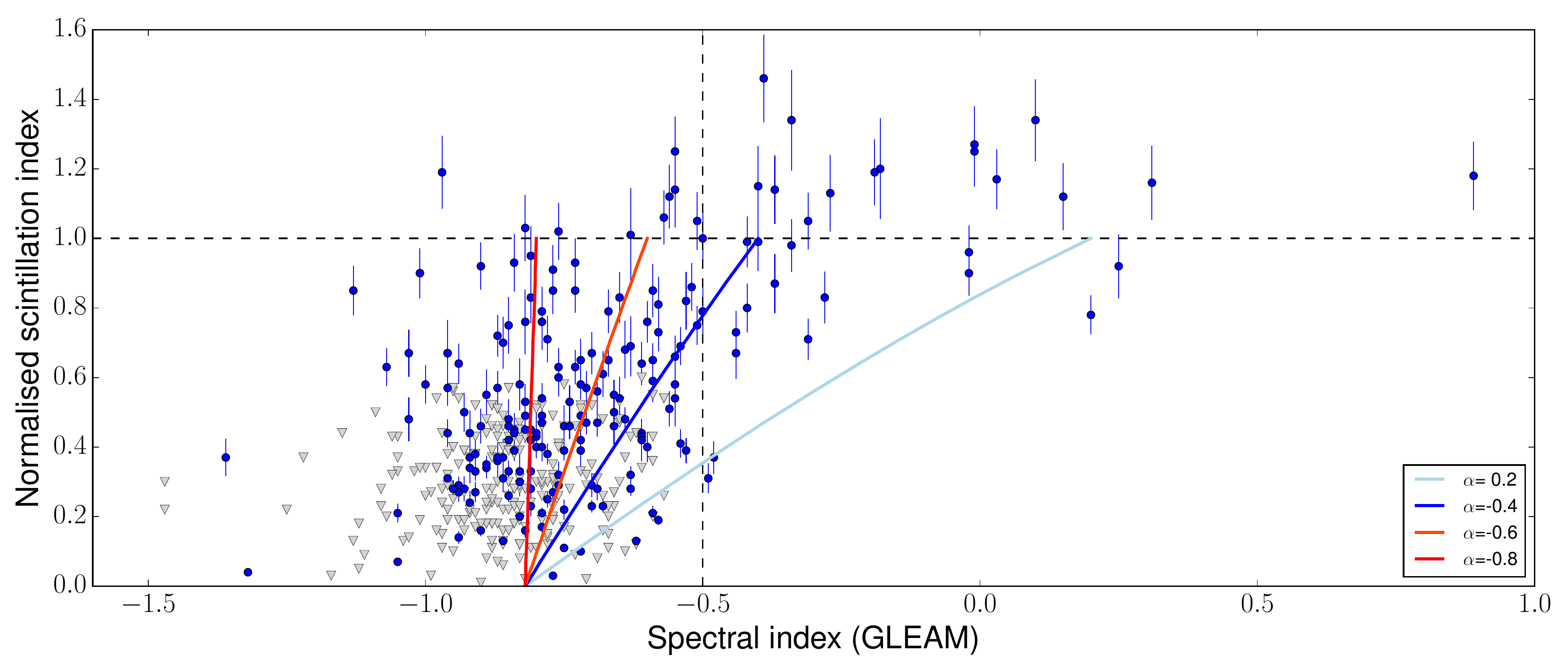}
\caption{We compare models to show scintillation indices for sources with varying degrees of dominance of the total flux by the compact components (high S/N subsample). The triangles represent the upper limits of normalised scintillation indices for non-scintillating sources. Models are calculated for compact components with four different spectral indices, shown with different coloured solid lines. It is noteworthy that the majority of composite sources appear to be due to compact components with steeper spectrum (e.g. hot-spots rather than flat spectrum cores in lobes). X-axis on the left hand side is truncated to show the main effects, removing two outliers from this plot that are described in Section \ref{Sec:ExtremeSteepSpecSrc}. }
\label{Fig:avg-nonscintillatingAll}
\end{figure*}
\end{center}

\subsubsection{Spectral index distribution between 162\,MHz and 1.4\,GHz}
We cross-matched all our sources against the 1.4\,GHz NVSS source catalogue \citep{Condon1998} to identify their higher-frequency counterparts.  We found matches for all but four of the 2550 sources in our field within a 60\,arcsec radius (two of these have counterparts just above 60\,arcsec radius and have been included as matches. The remaining two objects without NVSS counterparts are both non-scintillators). The angular resolution of NVSS images is 45\,arcsec (full width at half maximum), a factor of $\sim$ 3 higher than the MWA at 162 MHz, so it may resolve confused or extended radio sources into two or more components.  For all of our sources, the number of counterparts we found in our search are listed in Column 9 Table \ref{Tab:MainTable}.  For simplicity we have not attempted to interpret the multiple component sources since we don't know which component would be scintillating.  They are excluded from the following analysis. We calculated the spectral index for each source between 162 MHz, using flux densities from GLEAM, and 1.4 GHz NVSS flux densities. 

\begin{figure*}
\subfloat[Strong scintillator, steep spectrum]{\includegraphics[width = 3.2in]{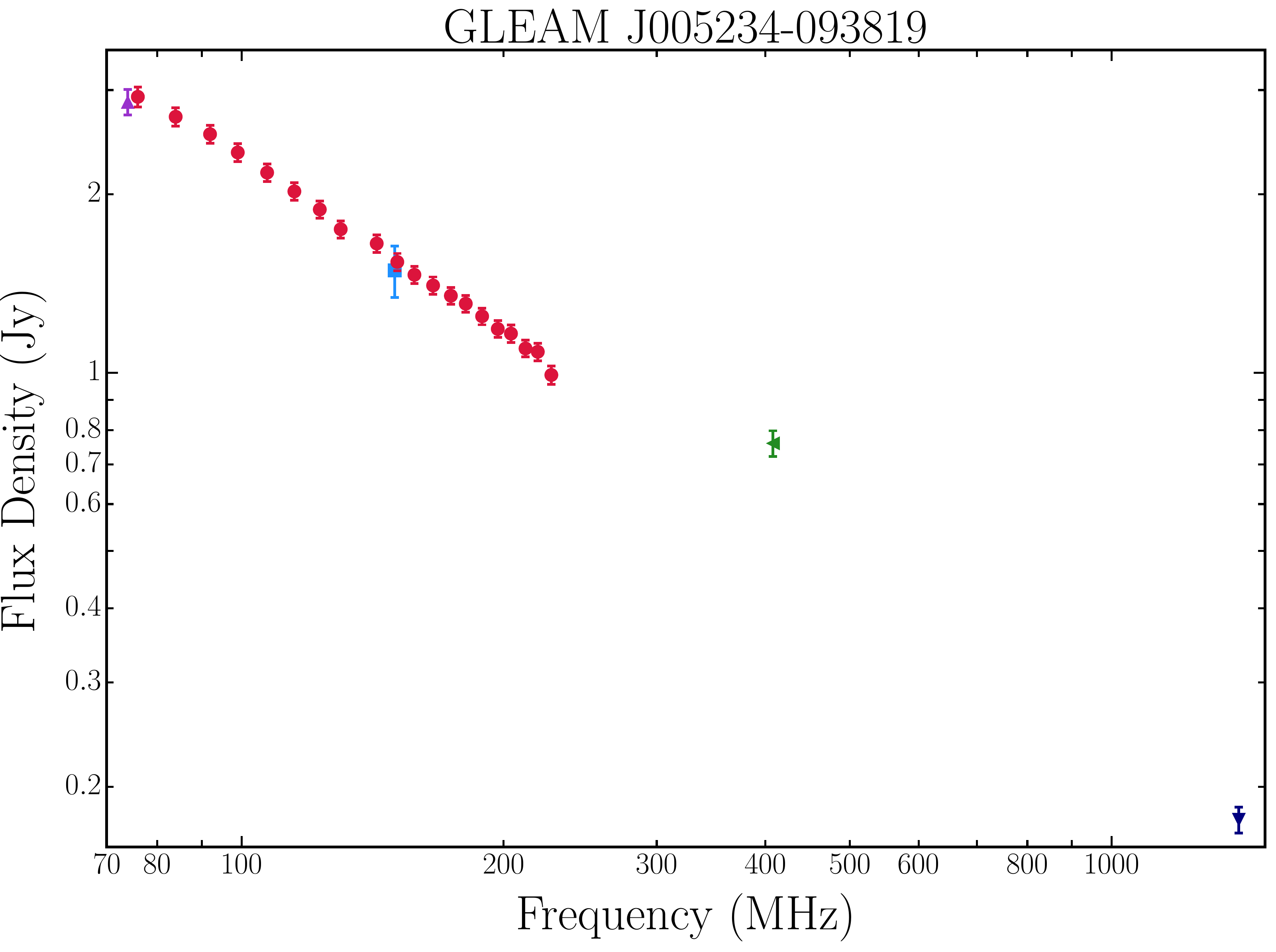}} 
\subfloat[Strong scintillator, peaked spectrum]{\includegraphics[width = 3.2in]{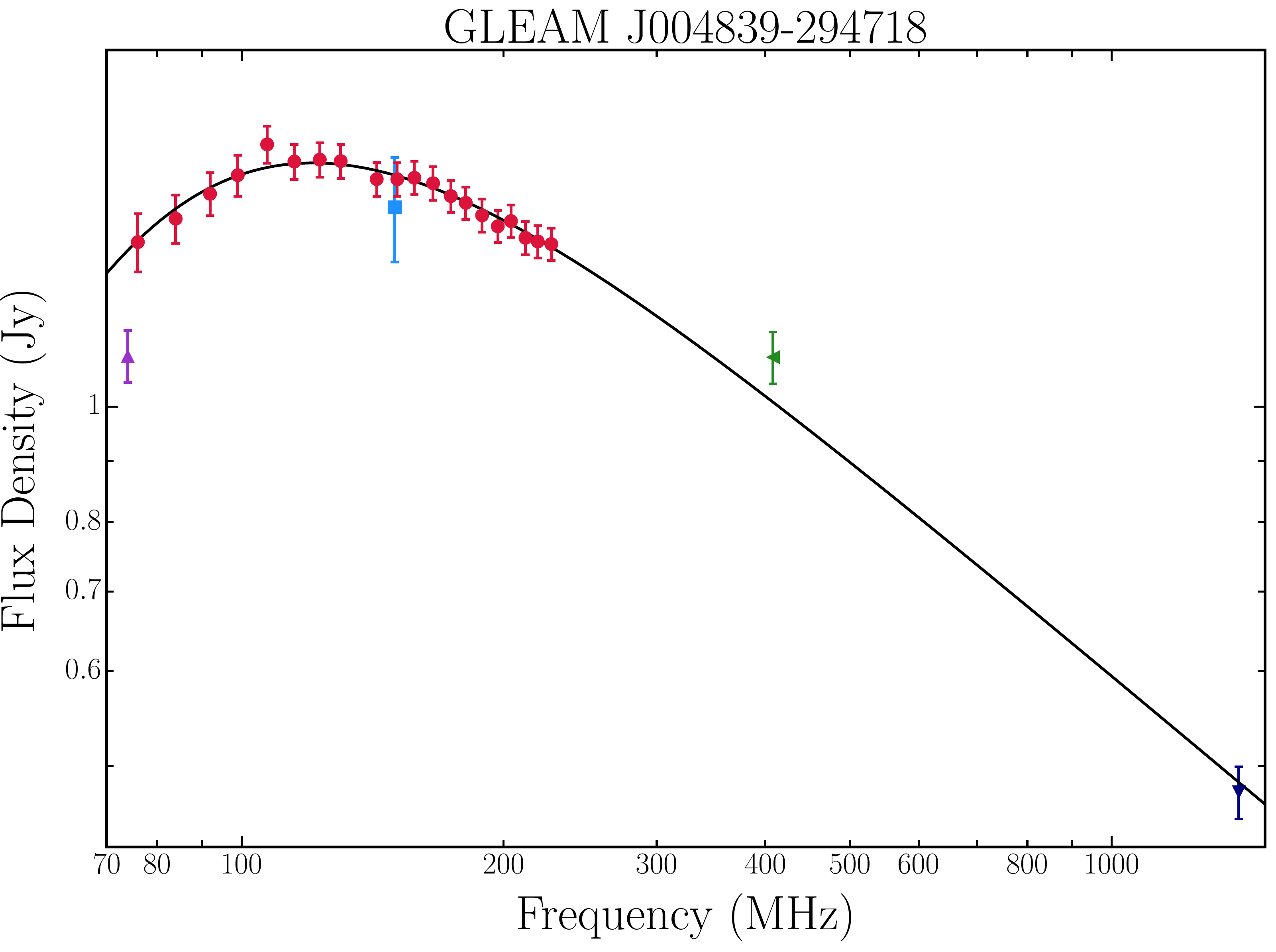}}\\
\subfloat[Moderate scintillator (embedded), steep \& flat spectrum]{\includegraphics[width = 3.2in]{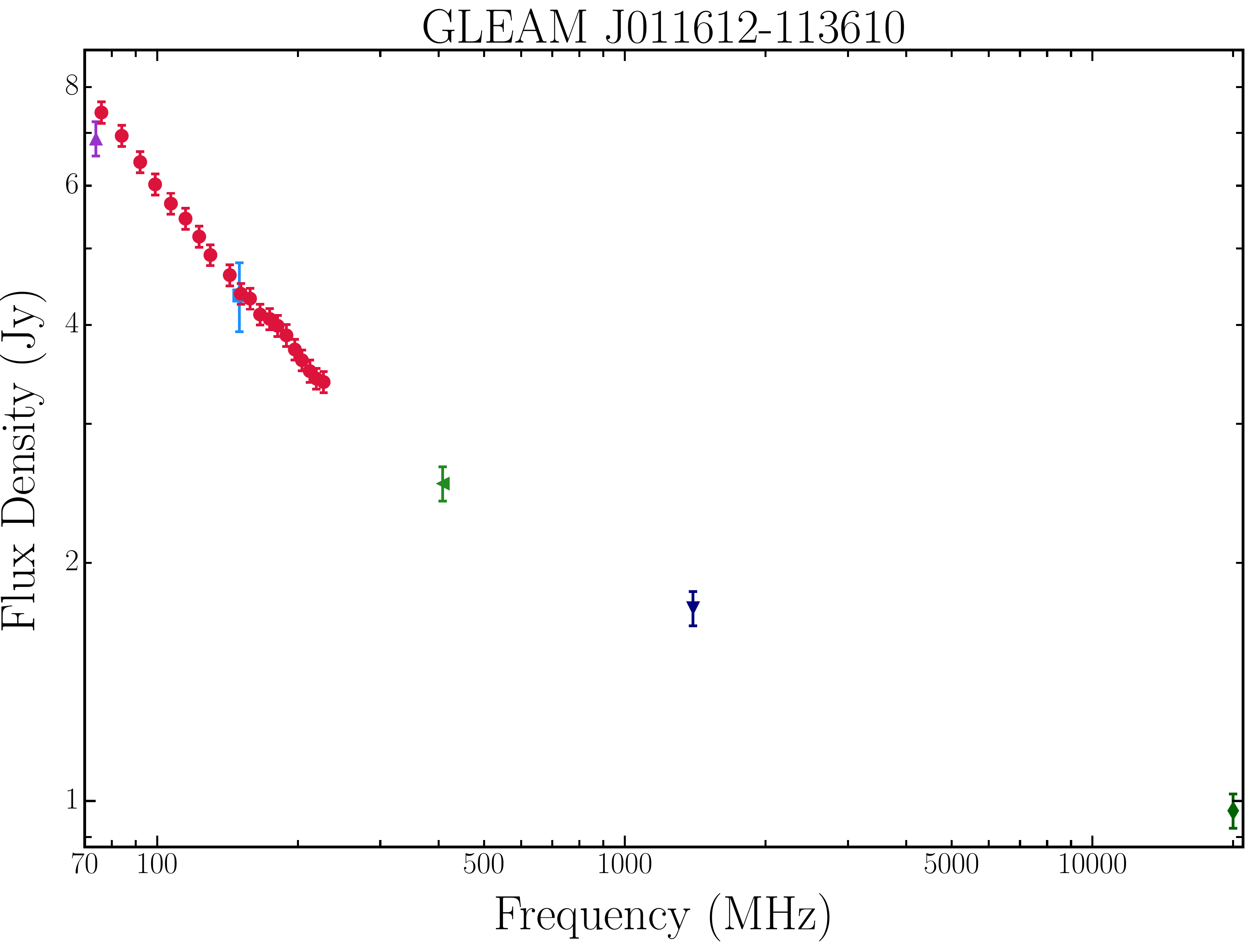}} 
\subfloat[Non-scintillator - steep spectrum]{\includegraphics[width = 3.2in]{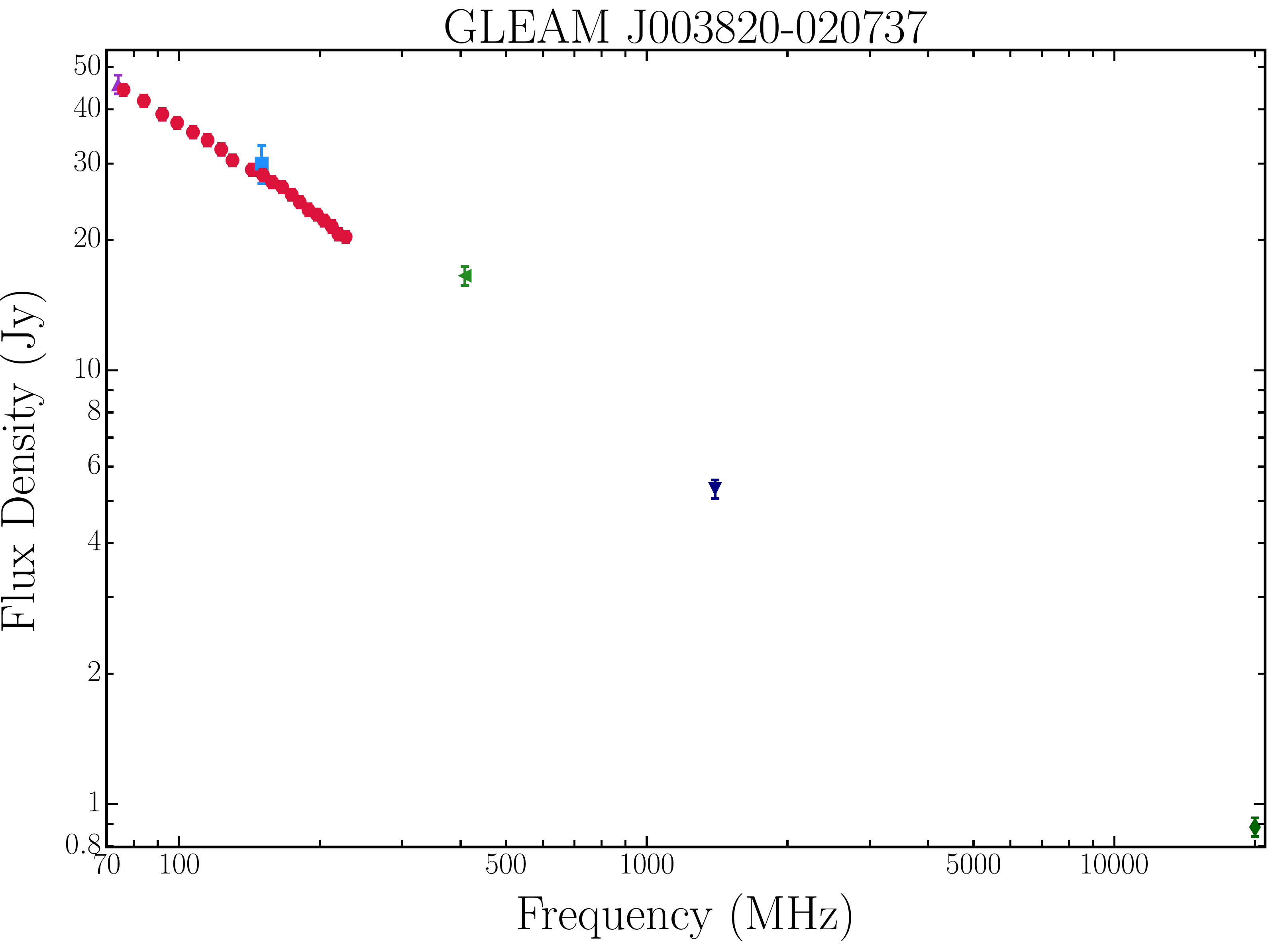}} 
\caption{Example SEDs of different types of objects identified using IPS and their low frequency spectra from GLEAM, and extended to higher frequencies. Flux density values have been compiled from different surveys. Red filled circles are from GLEAM, purple filled upward facing triangles are from VLSSr, blue filled squares are from TGSS ADR, green filled left facing triangles are from MRC, black filled downward facing triangles are from NVSS, and green filled circles are from AT20G survey.}
\label{Fig:exampleSEDs}
\end{figure*}

Representative SEDs of the different populations discussed in section \ref{Sec:AvgSED}, extended up to 1.4 GHz (and to 20 GHz, when available) are presented in Figure \ref{Fig:exampleSEDs}.

\begin{center}
\begin{figure}
\includegraphics[scale=0.4, angle=0]{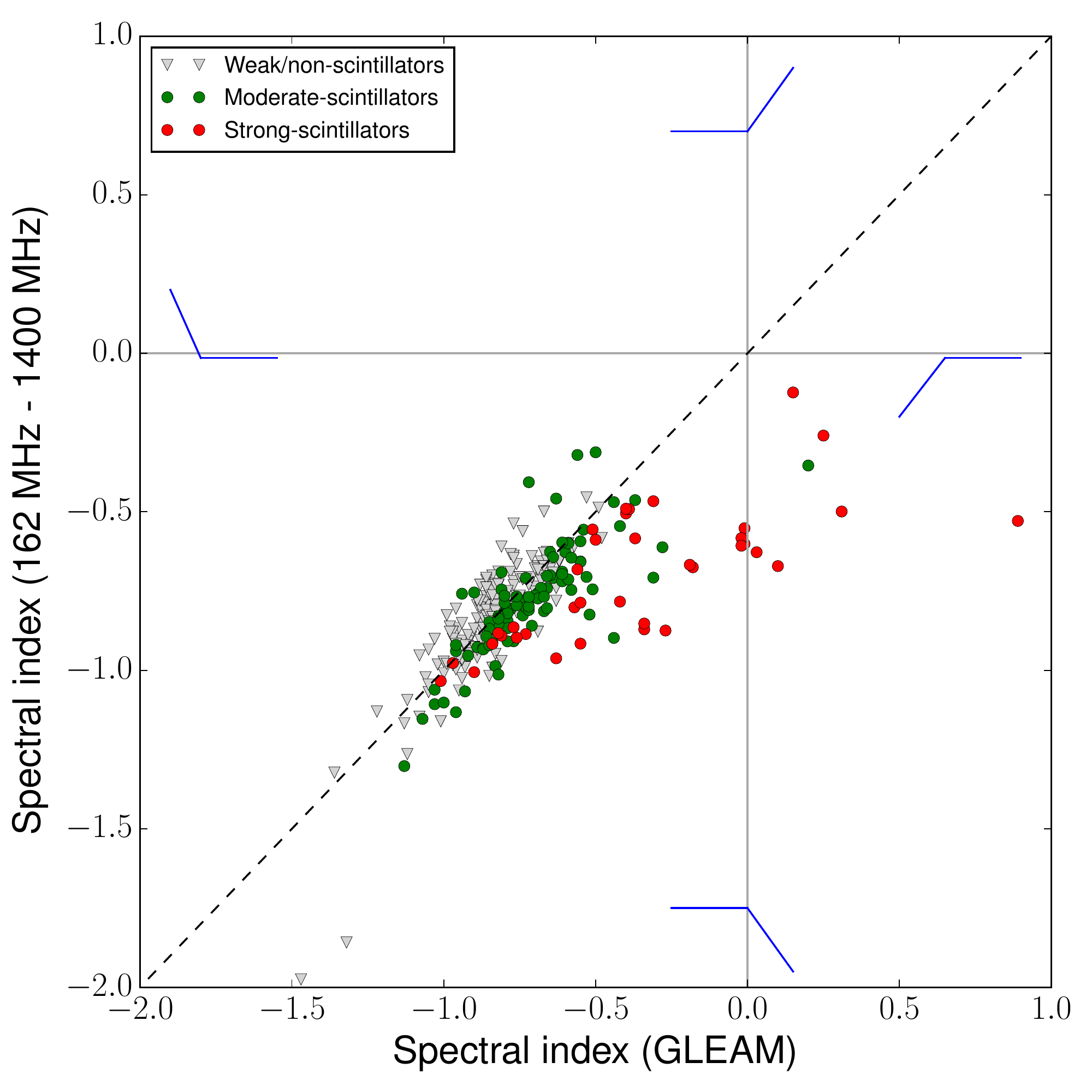}
\caption{Colour-colour plot to show the higher frequency spectral behaviour of strong, moderate and weak/non-scintillators (high S/N subsample). Strong, moderate, and weak/non scintillators are represented by red filled circles, green filled circles, and open triangles respectively. Plot is truncated at lower ends of both X and Y axes to make the plot more readable, removing an outlier that is described in Section \ref{Sec:ExtremeSteepSpecSrc}.}
\label{Fig:alphaDist_alt}
\end{figure}
\end{center}

The median spectral indices between 162 MHz and 1.4 GHz of strong scintillators, moderate scintillators, and weak/non scintillators are -0.68, -0.77, and -0.81 respectively (high S/N subsample). Compared to their GLEAM spectral indices, we find that the spectral index between 162 MHz and 1400 MHz of weak/non-scintillators becomes slightly less steep. This could be due to emergence of flat-spectrum cores unobservable at 162 MHz. For both moderately, and strongly scintillating sources, the overall spectral indices steepens at higher frequencies and this steepening of spectra becomes more pronounced  for the strong scintillators. This is evident in radio colour-colour plot (Figure \ref{Fig:alphaDist_alt}) which shows the low to high frequency SED trend of the the different populations. This steepening is observed because the high scintillators that we detect are mostly peaked-spectrum sources. This behaviour is quite different to what is seen at high frequencies (e.g. at 20 GHz) where the population of sub arcsecond sources are mostly flat-spectrum objects (as is seen in Figure \ref{Fig:ips_Fermi_at20g}). Thus, we find that the sub arcsecond source population detected is a function of frequency, and low frequency observations are specially important for identifying the rare peaked-spectrum objects in larger fractions, possibly from high redshifts as discussed below. 

\subsection{The partially scintillating sources - hot spots, jets, and cores}
\label{Sec:PartiallyScintillatingSources}
Three components of an extragalactic radio source may be small enough to scintillate: the core, the jets, and the hotspots. The degree of scintillation will also depend on the angular size and, hence, distance to the source.  For all three components, synchrotron self absorption at these low frequencies will put a limit on the smallest linear sizes that can be seen unless the rest frame has been redshifted to lower frequencies. For example a typical 1 kpc peaked spectrum source \citep[e.g.][]{Coppejans2016} with a peak at 1 GHz in the rest frame will will have its peak shifted into our band for a redshift of 5. Such a source will scintillate strongly, and will have a spectral index in our band which becomes steeper for high redshift sources.  

The models shown in Figure \ref{Fig:avg-nonscintillatingAll} are the scintillation index and average spectral index expected for the sum of a compact (scintillating) component and a non-scintillating lobe for a continuous range of the ratio of compact to extended components.  The dark blue, and light blue lines correspond to the case of a typical core with a range of spectral indices between -0.4, and +0.2, and the red lines a compact hot spot or jet with spectral indices between -0.8 and -0.6.  

Hotspots in radio lobes have linear sizes between 1 and 10\,kpc  \citep[e.g. ][]{Hardcastle1998}, with an average size of 2.5 kpc but the boundary between hotspots and lobes is not always well defined.  An average 2.5 kpc hotspot will have 50 percent scintillation at redshifts $>$ 0.4 (Figure \ref{Fig:scint-Z}).  However the scintillation index for a source with two hotspots will be reduced by a factor of $\sqrt2$ (slightly dependent on the position angle of the double with respect to the solar wind direction) and this will be further reduced by the flux density of the non-scintillating lobe. The majority of partially scintillating sources lie in the region of Figure \ref{Fig:avg-nonscintillatingAll} expected to be occupied by hot spots embedded in lobes.  While many hotspots have a spectral index similar to the radio lobes, \cite{Leahy1989} found spectral flattening in 30 percent of his sample of powerful QSOs and radio galaxies.  This result has been beautifully confirmed by the recent LOFAR images of Cygnus A.  \cite{McKean2016} showed that the Cygnus A hotspots have an inverted spectrum between 109 and 183MHz.  They also note that all explanations for the observed turn-over are problematic.  The Cygnus A hot spots have a linear size of 2 kpc \citep{Wright2004} so hotspots in sources like Cygnus A will have 10 percent scintillation for z $>$ 0.1 so they can also occupy the region in Figure \ref{Fig:avg-nonscintillatingAll} with partial scintillation and flatter spectra. This complicates the interpretation of the sources with partial scintillation and flatter spectral index since there will be a combination of weak flat spectrum cores and hotspots.  At redshifts higher than 1 the flat spectrum turnover will be redshifted out of the MWA band for a source like Cygnus A so only the relatively nearby hot spots will be confused with cores.  Note that the hotspots in Cygnus A are only about 1 percent of the total flux density so it would have very low scintillation index but if the hotspot to lobe ratio changes for sources at high redshift due to increased inverse Compton losses in the lobes as suggested by Ghisellini et al (2015) then these steeper spectrum weak scintillators could be another indicator of high redshift.

Sources with 1-D jets which are unresolved (less than the Fresnel scale) in cross section will scintillate but the degree of scintillation will be decreased by the square root of the length of the jet in Fresnel scale lengths.  \cite{Hardcastle1998} analysed the jet structure in a sample of FRII radio galaxies and found a median length of 20" so such jets would only be very weak scintillators.  Jets on smaller milli arcsec VLBI scales would be indistinguishable from the core sources.

The unresolved flat spectrum core sources (scintillation index $\approx$ 1) outnumber the sources in the core plus lobe region in Figure \ref{Fig:avg-nonscintillatingAll} which might be expected if there is a blazar population with beamed cores that are boosted and are out-shining the lobes.

\subsection{Variability of scintillating sources}
\label{sec:variability}
\noindent 
The GLEAM survey observations of our IPS field were made in November 2013. Our 162 MHz observations are also close in frequency to the 150 MHz observations of the TIFR GMRT Sky Survey (TGSS) project made between April 2010 and Nov 2011\footnote{GMRT online data archive\\ https://naps.ncra.tifr.res.in/goa/mt/search/basicSearch} with mean epoch of observation of 11 Jan 2011. With the recent release of the TGSS alternative data release \citep[TGSS ADR,][]{Intema2017}, we can search for variations in flux density between the GLEAM 151 MHz and the TGSS ADR\,150 MHz observations across a time interval of 1 to 3 years.
A similar technique was used by \cite{Murphy2017a}\ to search for low-frequency radio transients across the full region of overlap between TGSS ADR and GLEAM. 

The $\sim$ 25\,arcsec angular resolution of TGSS ADR may resolve out some flux from extended sources. However our IPS dataset can be used to identify  high-scintillation (sub arcsecond compact) objects making us uniquely placed to identify variable compact sources at low radio frequencies. 
We selected strong scintillators (NSI$\geq$0.90) from the high S/N sample  (37 objects), and found TGSS counterparts for all these sources within 30 arcsec of their GLEAM positions. A comparison of flux density between GLEAM 151 MHz and TGSS ADR suggests that the TGSS flux density in our IPS field is scaled below the GLEAM flux density by a factor of $\sim$ 1.15 (N. Hurley-Walker private communications, see also  \cite{Murphy2017a} and \cite{Hurley-Walker2017b} for more detailed discussions of the TGSS and GLEAM flux-density scales). We therefore corrected the flux densities for this relative offset before searching for variability. 

To check for variability, we measured a debiased variability index for each source as outlined by \cite{Sadler2006}. This index takes into account the uncertainties in individual flux densities measurements to determine whether a source shows significant variability. 
For 36 of the 37 sources in our sample there was no significant variability between the two epochs, with an upper limit of 8 per cent or less for the debiased variability index. 
The only object that showed significant variability was the peaked-spectrum source GLEAM J013243-165444. This is a known blazar (at redshift z=1.02) with a \fermi\ gamma-ray detection, and is discussed in more detail in Section~\ref{sec:ips_fermi} below. Thus, we find $\sim$ 2.6 percent of compact objects at 162 MHz appear to show strong variability over the 2-3 year time period.

As a practical outcome of the IPS work we can estimate the space density of the compact sources. At the centre of the field of our observation we detect approximately one scintillating source per square degree of sky, a detection rate that is similar to Paper I. As discussed above these compact sources show low variability, and have known compactness at sub arcsecond scales making them excellent candidate calibrators for future low frequency radio telescopes. This is important for SKA-low as its ionospheric calibration strategies require the knowledge of space density of compact objects at low frequencies.

\subsection{Comparison with \fermi\ gamma-ray sources }

\subsubsection{Low-frequency properties of \fermi\ blazars}
\begin{center}
\begin{figure}
\hspace*{-1.0cm}
\includegraphics[scale=0.78, angle=0]{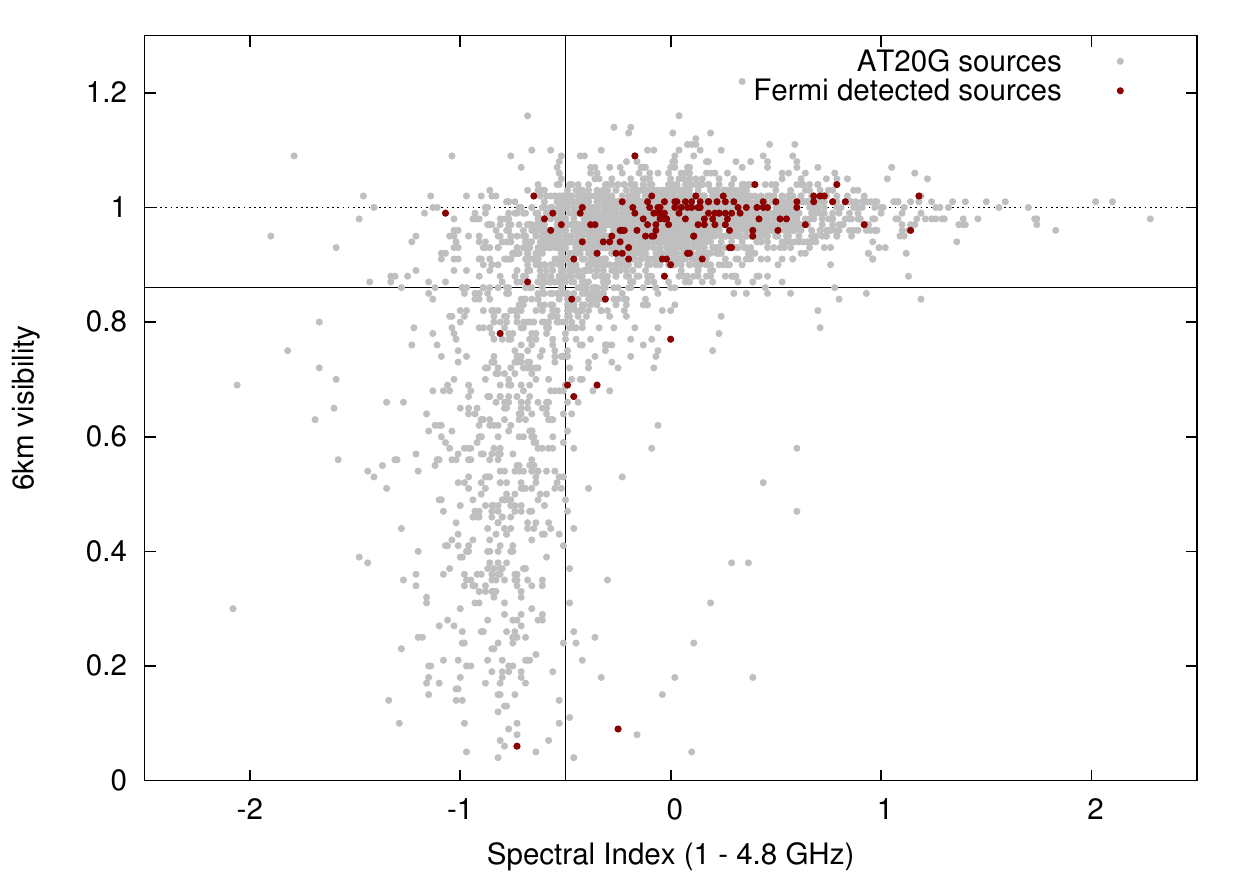}
\caption{High-frequency radio properties of \fermi\ blazars detected in the Australia Telescope 20\,GHz (AT20G) survey (\protect\cite{Murphy2010}, figure is from \protect\cite{Chhetri2013b}). Grey and red filled circles represent AT20G sources, and AT20G sources with \fermi\ counterparts respectively. As explained by \protect\cite{Chhetri2013}, sources with 6-km visibility larger than 0.86 are expected to have angular sizes smaller than about 0.2\,arcsec at 20 GHz. The only \fermi\ sources in the AT20G sample with significantly-extended radio emission at 20 GHz are the nearby starburst galaxies NGC\,253 and NGC\,4945. }
\label{Fig:ips_Fermi_at20g}
\end{figure}
\end{center}

The vast majority (98\%) of radio-loud AGN detected as gamma-ray sources by \fermi\ belong to the subclass of blazars, where the jet axis is closely aligned with our line of sight suggesting relativistic beaming \citep[e.g.][]{Giroletti2016a}. As can be seen from Figure \ref{Fig:ips_Fermi_at20g}, their high-frequency radio emission is usually completely dominated by a compact radio core.

Most blazars are flat-spectrum radio sources, and remain poorly studied at low radio frequencies. \cite{Giroletti2016b} cross-correlated the 6100 deg$^2$ Murchison Widefield Array Commissioning Survey catalogue \citep[MWACS:][]{Hurley-Walker2014} with several blazar catalogues, and found low-frequency radio counterparts for 79 out of 174 (45\%) \fermi\ gamma-ray blazars in the MWACS survey area. They found no correlation between the gamma-ray energy flux of these objects and the MWA flux density at 180\,MHz, indicating that non-beamed lobes may dominate the radio flux density of \fermi\ blazars at MWA frequencies. We can now test this using the additional structure information from our IPS observations at 162\,MHz. 

\subsubsection{\fermi\ sources in the MWA IPS field}
We used the \fermi\ large area telescope (LAT) 4 year point source gamma ray catalogue \citep[3FGL;][]{Acero2015} to identify gamma-ray sources within the field of view of our IPS observation. There are 83 \fermi\ sources in the field: 71 blazars (30 BL Lacs, 24 flat-spectrum radio quasars, or FSRQ, and 17 blazar candidates of uncertain type, or BCU), 1 starburst galaxy (NGC\,253), 2 pulsars, and 9 `unassociated' gamma-ray sources (UGS) with no identified radio counterpart. 

To identify the low-frequency radio counterparts of these \fermi\ sources, we initially focussed on the known blazars, for which accurate positions are reported in the associated third AGN catalogue \citep[3LAC;][]{Ackermann2015}; we used a conservative 10\,arcsec uncertainty for the blazar position (typically known from NVSS, if not from VLBI) and 30\,arcsec uncertainty for the radio (GLEAM) positions for sources in our IPS field. 22 of the 71 \fermi\ blazars in the field (31\%) were matched with a source in the standard image. 
The remaining \fermi\ blazars have a radio counterpart listed in the 3LAC catalogue, but they are too faint at 162\,MHz to be detected in our standard image. 

We further searched within the 95\% gamma-ray confidence ellipse for the 12 remaining gamma-ray sources in our field, still assuming a 30\,arcsec uncertainty for the IPS source positions. We found two more matches: the starburst galaxy NGC\,253 and one unassociated \fermi\ source with a candidate MWA counterpart (GLEAM J003039-232300). 
At this stage it is unclear whether this is a genuine radio association with the \fermi\ source 3FGL J0031.2-2320 and follow-up imaging at higher radio frequencies is needed to investigate this further. 
Table \ref{tab:fermi_stats} summarises the low-frequency radio properties of \fermi\ sources in our field. 

\begin{center}
\begin{figure*}
\includegraphics[scale=0.44, angle=0]{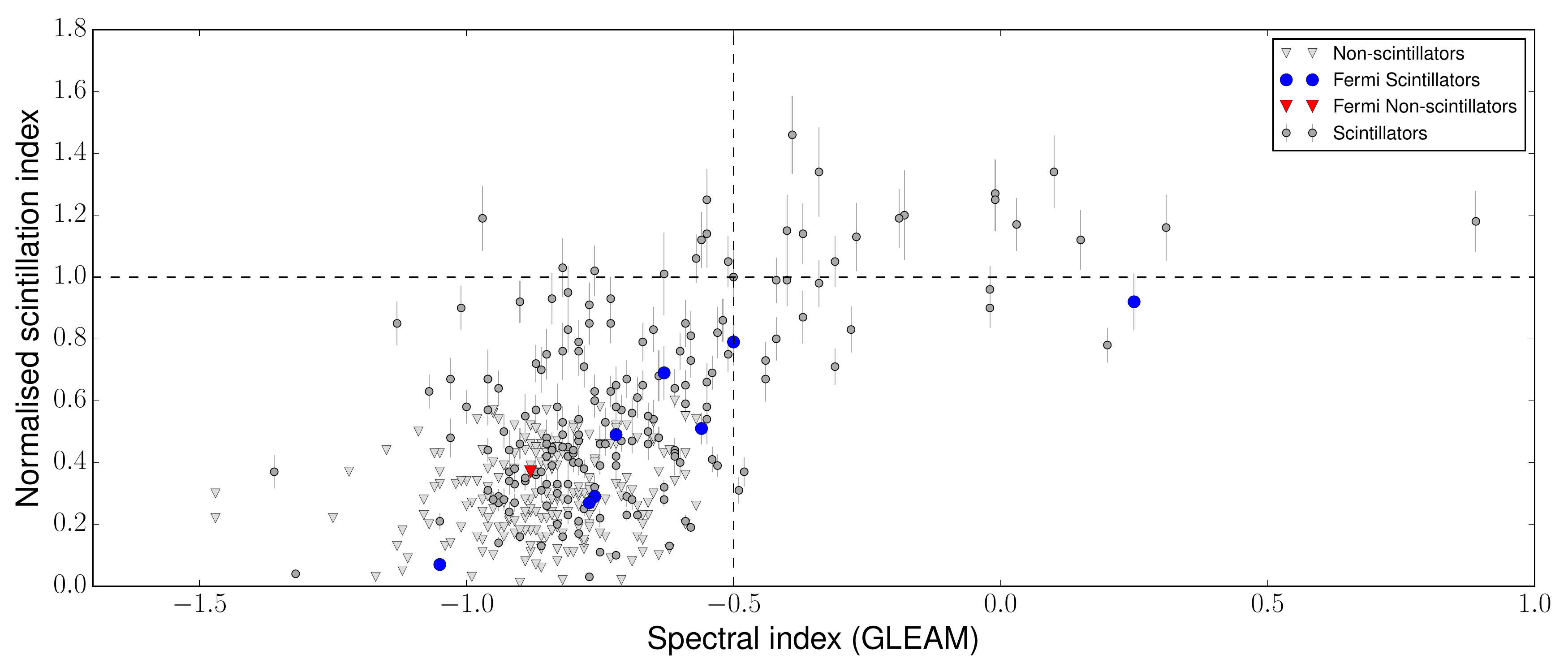}
\caption{Plot of normalised scintillation index versus GLEAM spectral index for the high S/N subsample. The \fermi-detected MWA sources are overplotted as blue and red points for scintillating and non-scintillating sources respectively. The X-axis on the left hand side is truncated to show the main effects, removing two outliers from this plot (described in Section \ref{Sec:ExtremeSteepSpecSrc}). }
\label{Fig:ips_Fermi_alpha}
\end{figure*}
\end{center}

\begin{table}
\resizebox{\columnwidth}{!}{%
\begin{tabular}{lcccrrr}
\hline            
 & \multicolumn{3}{c}{\fermi\ sources in the MWA field} \\
\multicolumn{1}{l}{Type} & \multicolumn{1}{c}{Detected at} &      \multicolumn{1}{c}{Non-detection} & \multicolumn{1}{c}{Radio}  & High S/N \\     
   & 162\,MHz &at 162\,MHz & det. rate & ($\geq$12.5) \\
\hline
FSRQ      & 12 & 12 &  50\% &  6 \\
BL Lac    & 6 & 24 &  20\%  &  1 \\
BCU       & 4 & 13 &  24\%  &  1 \\
Starburst & 1 & 0  & 100\%  &  1 \\
Pulsar    & 0 & 2  & 0\%    &  .. \\
UGS       & 1 & 8  &  11\%  &  .. \\
&&&&& \\
Total     & 24  & 59 & 29\%  &  9 \\
\hline
\end{tabular}
}
\caption{Summary of the 162\,MHz radio properties of the 83 \fermi\ gamma-ray sources in our MWA field. FSRQ = Flat-spectrum radio QSO; BCU = blazar of unknown class; UGS = Unidentified gamma-ray source. }
\label{tab:fermi_stats}
\end{table}

\subsubsection{IPS properties of \fermi\ sources}
\label{sec:ips_fermi}
Figure \ref{Fig:ips_Fermi_alpha} plots the normalised scintillation index of all our radio-matched \fermi\ sources with high S/N against the radio spectral index within the 72-231\,MHz GLEAM band. In contrast to the \fermi\ blazar population observed at 20\,GHz (Figure \ref{Fig:ips_Fermi_at20g}), many of the 162\,MHz \fermi\ blazars have steep radio spectra and their scintillation properties are consistent with a compact component embedded within more extended low-frequency emission. 

Nine of the 24 \fermi\ matches with 162\,MHz sources are in the high S/N subsample and can be split into scintillation classes in a reliable way (see Section~\ref{sec:ngc253}). One of these objects is the starburst galaxy NGC\,253. The median normalised scintillation index for the remaining eight objects is 0.51, and only one source (the FSRQ GLEAM J013243-165444), already identified in Section~\ref{sec:variability} as a variable radio source, shows strong scintillation with NSI $\geq$ 0.9. 
Within our full high S/N subsample, the gamma-ray detection rate is only marginally higher (5/134, or 3.7$\pm1.7$\%) for objects with moderate or strong scintillation (NSI $\geq$ 0.4) than for objects with weak or no scintillation (4/221, or 1.8$\pm0.9$\%), and the difference is not significant in this relatively small sample.

\begin{figure*}
\subfloat{\includegraphics[width = 3.2in]{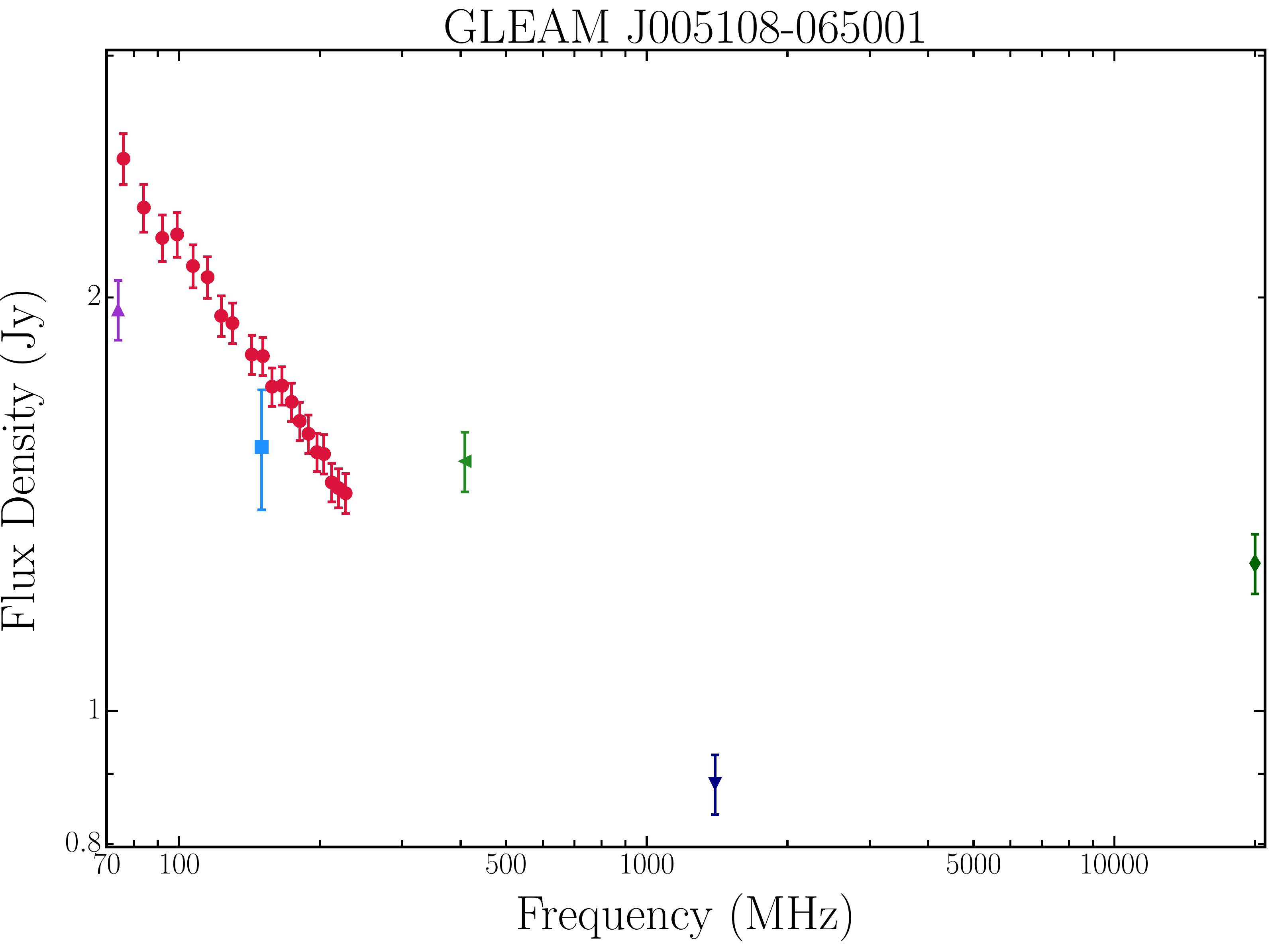}} 
\subfloat{\includegraphics[width = 3.2in]{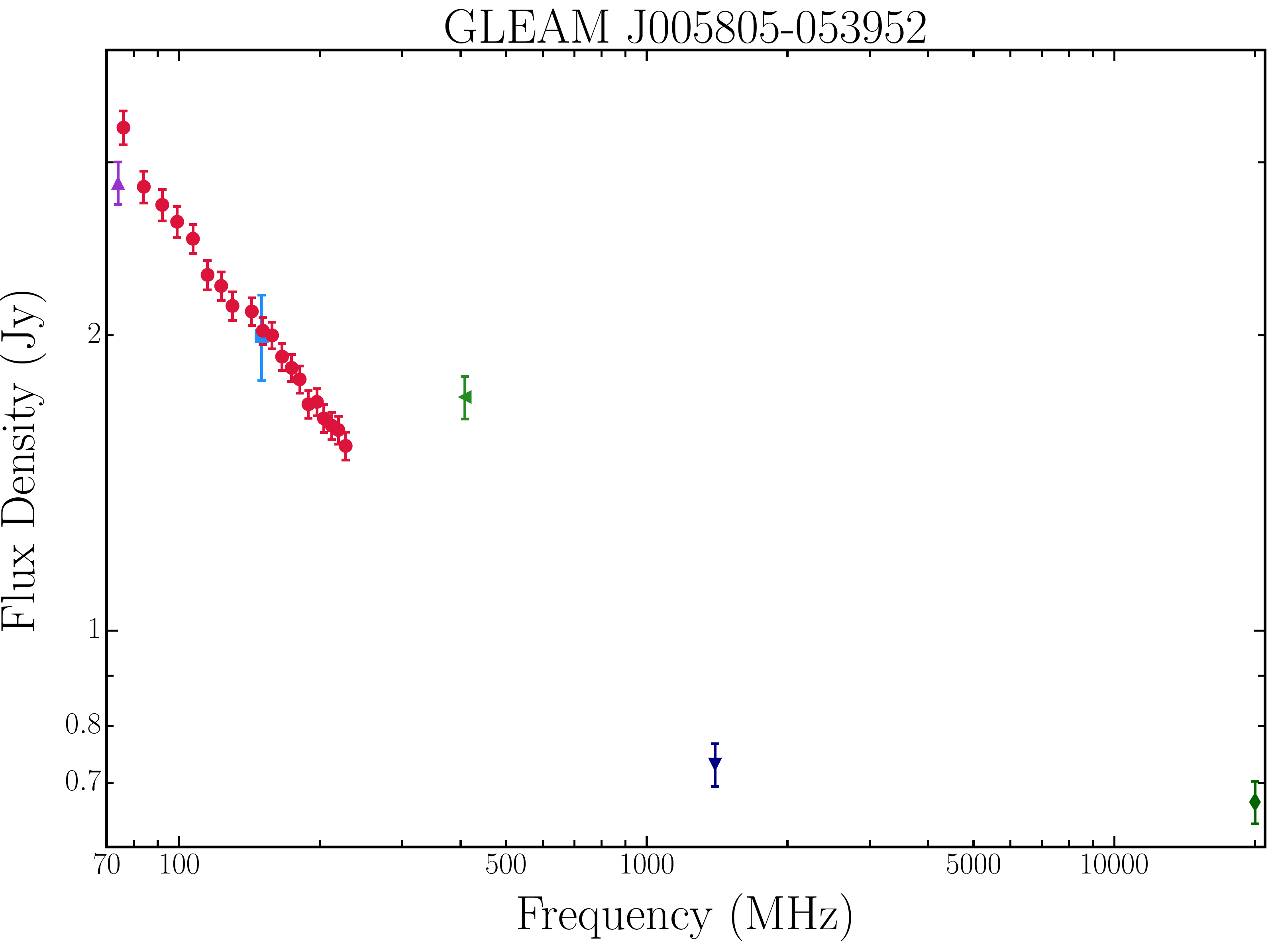}}\\
\subfloat{\includegraphics[width = 3.2in]{GLEAM_J011612-113610}}
\subfloat{\includegraphics[width = 3.3in]{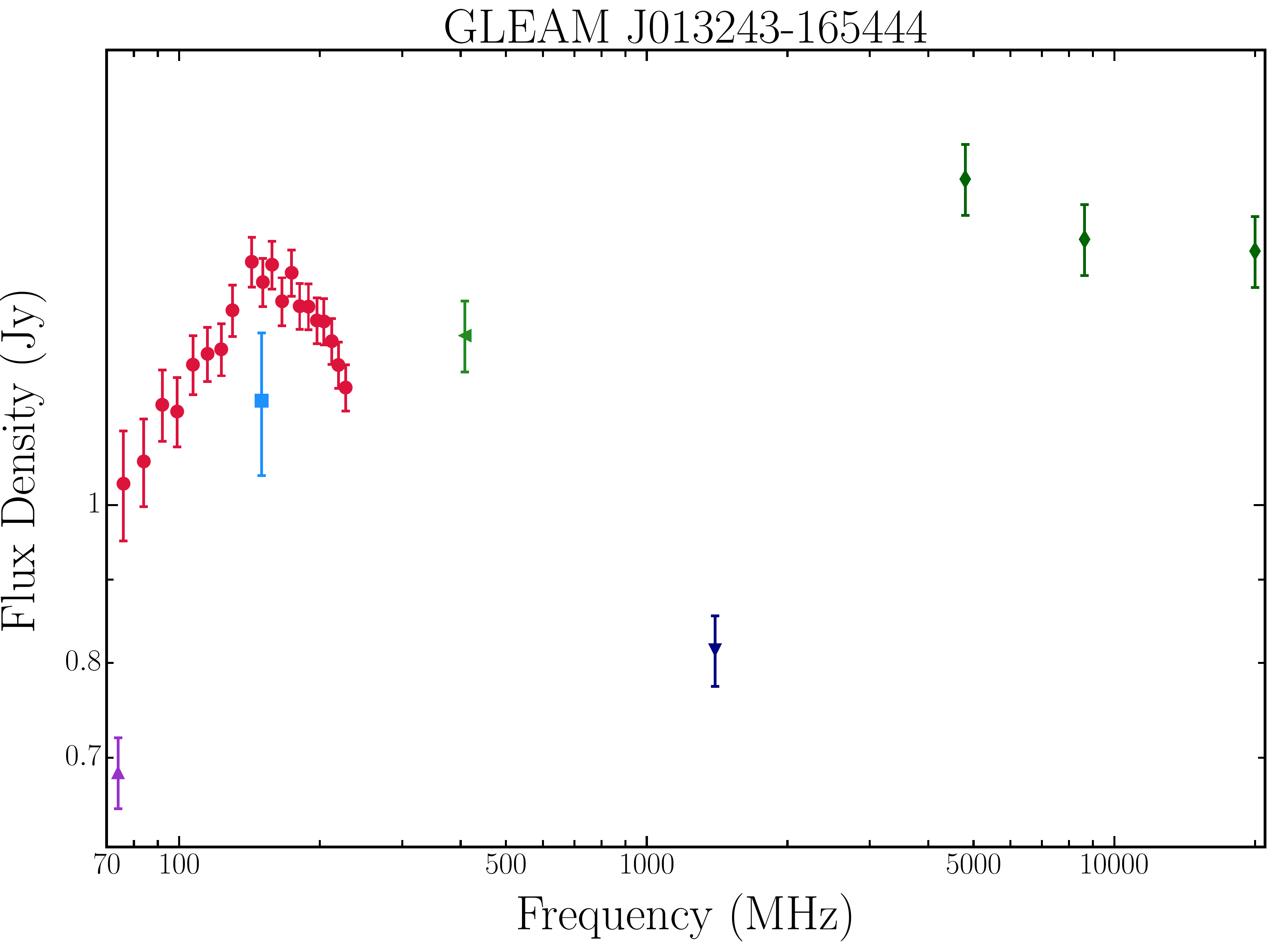}} 
\caption{Radio SEDs of four scintillating MWA sources that have \fermi\ gamma-ray counterparts and are also detected by the AT20G survey at 20\,GHz. All four of these objects are listed as flat-spectrum radio quasars in 3FGL, and three of them show moderate levels of scintillation at 162\,MHz.  The peaked-spectrum source GLEAM J013243-165444 is the only strongly-scintillating source and is discussed in more detail in Section \ref{Sec:J013243-16544}. Flux density values have been compiled from different surveys. Red filled circles are from GLEAM, purple filled upward facing triangles are from VLSSr, blue filled squares are from TGSS ADR, green filled left facing triangles are from MRC, black filled downward facing triangles are from NVSS, and green filled circles are from AT20G survey. }
\label{Fig:ips_Fermi}
\end{figure*}

20 of the 22 \fermi\ blazars detected at 162\,MHz are also detected in the AT20G survey at 20\,GHz (all classified as \fermi\ blazars in the 3FGL catalogue). Almost all of these are very compact sources at high frequency (17/20 have a 6-km visibility $\geq$ 0.86 in Figure \ref{Fig:ips_Fermi_at20g}). With the exception of GLEAM J013243-165444, a QSO at redshift $z=1.02$ \citep{Wright1983} discussed above and in Section~\ref{Sec:J013243-16544}), none of the AT20G-Fermi sources in our high S/N sample are strong scintillators (with NSI $\geq$ 0.9) at 162\,MHz.
The radio SEDs of four of the scintillating sources also detected at 20\,GHz are shown in Figure \ref{Fig:ips_Fermi}. In all four cases the measured 20\,GHz flux density lies well above the extrapolation of the low-frequency radio spectrum from GLEAM, showing the emergence of the blazar core at high frequencies. 

Overall, our IPS results are consistent with a picture in which many \fermi\ blazars are compact radio sources at high  frequencies, but are embedded in more extended structures that dominate the low-frequency emission. As a result, the compactness of a source at 162\,MHz is not a reliable indicator of whether it hosts a blazar core. 

Our observations were only sensitive enough to detect about 30\% of the known radio counterparts of \fermi\ blazars, so the current sample may over-represent blazars with extended low-frequency emission -- simply because such objects are more likely to be bright enough at 162\,MHz to lie above our detection limit. We can however make a statistical statement about the overall properties of {\it all}\ \fermi\ blazars in our field. The field contains 54 blazars of known class (FSRQs and BL Lacs), 18 of which are detected at 162\,MHz (see Table \ref{tab:fermi_stats}). Of these, only two are strong scintillators and 10 have NSI $<$ 0.9 (six have non-restrictive upper limits in NSI). Thus we can already set a lower limit of 19\% for the fraction of all \fermi\ blazars that are likely to be embedded in extended low-frequency emission rather than being dominated by a sub-arcsec scale compact component at 162\,MHz.

\subsection{Comparison with the Roma-BZCat blazar catalogue}
We used v5.0 of the Roma-BZCat of all known blazars \citep{Massaro2015} to compile a list of 190 blazars in the field of our observations and cross-matched this list with the sources in Table 1 assuming a positional uncertainty of 10\,arcsec for the BZCat coordinates (which typically come from NVSS) and 30\,arcsec for those from MWA. 48 of the 190 blazars (25\%) were matched with a 162\,MHz source, and 10 of these are in the high S/N subsample sample and can be divided into scintillation classes in a reliable way. 

Within the high S/N sample, the detection rate is again higher for strong or moderately scintillating sources with NSI $\geq$0.4 (6/134, or 4.5$\pm1.8$\%) than for sources with NSI $<$0.4 (4/221, or 1.4$\pm0.8$\%). The median normalised scintillation index for the Roma-BZCat blazars in the high S/N subsample is 0.48, i.e. similar to the median value of 0.51 for the \fermi\ sources in Section~\ref{sec:ips_fermi}.

\subsection{Scintillation properties of peaked-spectrum sources}
\label{Sec:PeakedSpectrumSources}
While we expect small diameter sources to become synchrotron self absorbed at low frequency resulting in a peaked spectrum, the very high fraction of scintillating sources that have radio spectra peaking below 1\,GHz is one of the most surprising results to emerge from our analysis.

\cite{Callinghametal17} recently published a catalogue of 1,483 sources selected from the GLEAM survey that display spectral peaks between 72\,MHz and 1.4\,GHz. 65 of these peaked-spectrum sources lie within our IPS field, and 21 of them are in the high S/N subsample. 
All 21 of these sources are moderate to strong scintillators with NSI $>$ 0.7, and 17/21 (81\%) have NSI $>$ 0.9. In other words, none of the well-observed peaked-spectrum sources has more than about 30\% of their flux density in extended lobes. 

\begin{center}
\begin{figure}
\includegraphics[scale=0.33, angle=0]{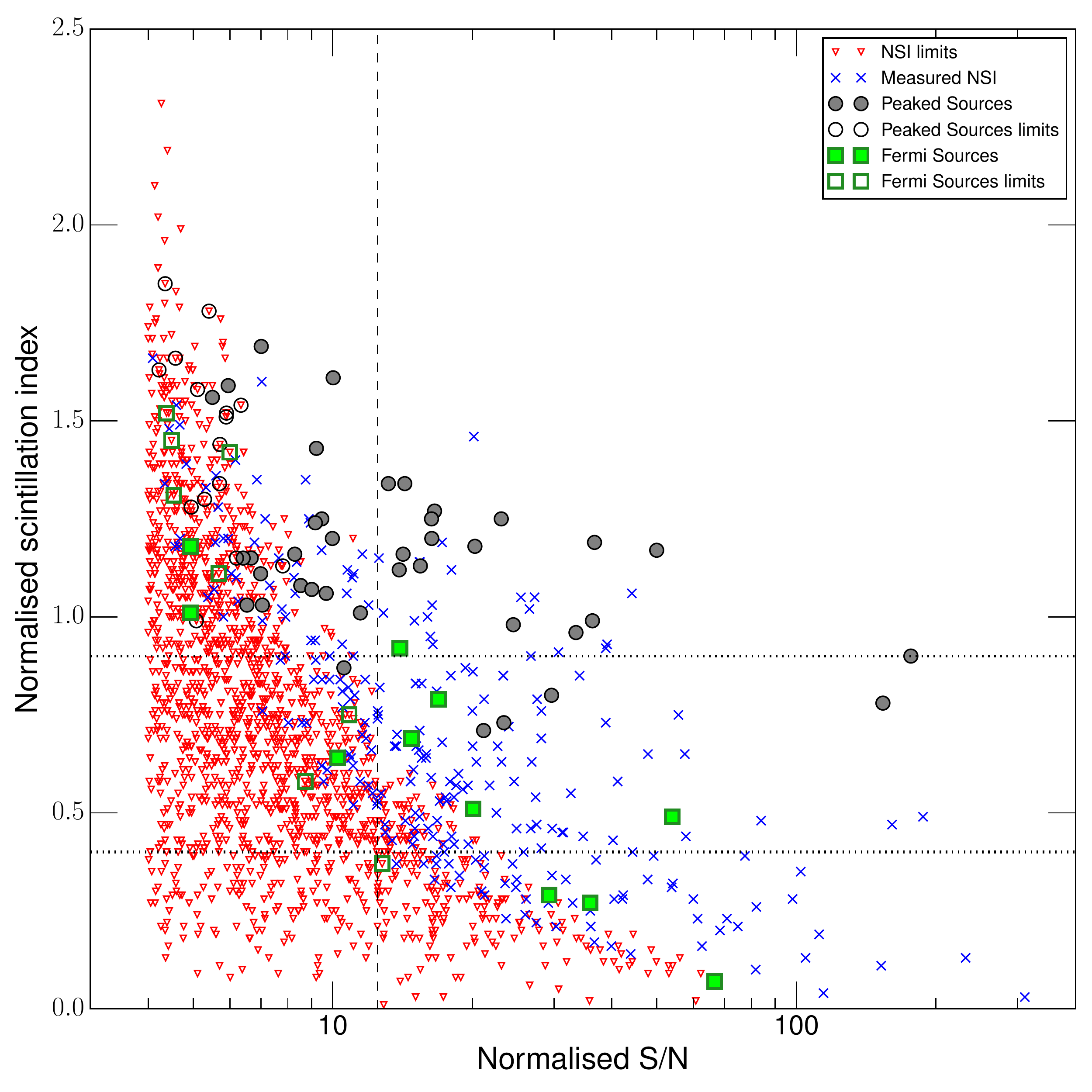}
\caption{Plot of normalised scintillation index (NSI) versus normalised S/N for sources listed in Table 1, with normalised S/N cut-off made at 4. Sources with detected IPS are shown in blue, and the 5-$\sigma$ upper limits for sources with no IPS detection in red. Peaked-spectrum sources from Callingham et al.\ (2017) are marked with black circles, and 3FGL \fermi\ sources from Acero et al.\ (2015) with green squares. Horizontal dotted lines delineate the classes of strongly-scintillating sources (NSI $\geq$ 0.9), moderate scintillators (0.4 $\geq$ NSI $<$ 0.9), and sources with weak or no scintillation (NSI $<$ 0.4). The vertical dashed line shows the normalised S/N = 12.5 limit applied for much of the analysis in this paper.}
\label{Fig:fermi_peaked}
\end{figure}
\end{center} 

For the 44 objects in the low S/N sample, 22 have upper limits on NSI which are not restrictive enough to allow us to make any statement about sources with moderate scintillation.
We can however say something about the strongly-scintillating population even here. There are 20 sources in the low S/N sample that have detected scintillation and NSI $\geq$ 0.9, and 2 with values or upper limits that have NSI $<$ 0.9. Thus the detection rate for strong scintillation remains just as high for these weaker peaked-spectrum sources (91\% at S/N $<$ 12.5 versus 81\% at S/N$\geq$ 12.5). 
  
Figure \ref{Fig:fermi_peaked} plots the normalised scintillation index (NSI) against normalised S/N for sources in Table 1, with normalised S/N cut-off made at 4.0 to show profile of NSI and meaningful limits only. Peaked-spectrum sources from \cite{Callinghametal17} are marked with black circles and the \fermi\ gamma-ray blazars with green squares. From this, we can see that the peaked-spectrum sources scintillate much more strongly than the \fermi\ blazars at 162\,MHz. Indeed, peaked-spectrum sources account for almost half (47\%) of the entire population of strongly-scintillating sources in our high S/N subsample. The one exception is the \fermi\ source J013243-165444 discussed in Section \ref{Sec:J013243-16544} which is a blazar and its spectral shape may well be determined by its current activity.  

Since 100\% of the peaked spectrum sources in this sample have high scintillation it is possible that all sources with high scintillation will have intrinsically peaked spectra but some may have their peak in the part of the spectrum not observed, i.e. a peak either below 70\,MHz or above a 1400\,MHz in the observed frame.  There are no strong scintillating sources in Figure \ref{Fig:alphaDist_alt} which could have a peak above 1400\,MHz so all the red points (strong scintillators) to the left of the vertical line at 0 are strong candidates for sources at high redshift with their spectral peak red-shifted to below 70\,MHz.

It is perhaps not surprising that most of the peaked-spectrum sources are dominated by compact components on sub-arcsec scales.  Observationally, we know that sources with spectral peaks at frequencies above $\sim$200\,MHz are typically less than 1\,arcsec in angular size \citep{Snellen2000}, while sources with spectral peaks below this are generally larger. Low-frequency VLBI observations of a small sample of sources with radio spectra peaking below 1\,GHz \citep{Coppejans2016} also show that these are compact sources less than about 1\,kpc in size and appear to be genuinely young radio galaxies. The power of our IPS technique is that it can probe angular resolutions similar to those imaged by low-frequency VLBI, but over a much larger field of view and for hundreds of objects simultaneously. 

\cite{Orienti2014} have recently extended the work of \cite{ODea1998} to derive an empirical relationship between peak frequency and linear size for peaked-spectrum radio sources across a wide range in frequency.  
Based on this empirical relation, we might expect to see some relationship between normalised scintillation index and spectral peak for the sources in our sample, with the caveat that in many cases the redshift of the source (and hence the rest-frame spectral peak) is currently unknown. A test of this kind is beyond the scope of the current paper, but would be interesting to carry out in future.

Finally, we note that the strong prevalence of IPS in the peaked-spectrum population is unsurprising if the peak frequency, peak flux density, and angular size follow the relation expected for a self-absorbed synchrotron source.

The absence of IPS in a given source may provide a strong constraint on its angular size which, under certain circumstances may invalidate the viability of synchrotron self-absorption as an explanation for its spectral behaviour below the peak of the emission. The assumption that the spectral turnover in a peaked-spectrum source is due to synchrotron self-absorption implies a relationship between the direct measurables size, flux density ($S_t$), redshift, and peak frequency ($\nu_t$) with the value of the magnetic field strength, $B$.  Specifically, a lower limit on source size of $\theta_t$ implies a lower limit to the product $B (1+z)$ \citep[e.g.][]{KellermanPaulinyToth81} of
\begin{eqnarray}
\left[ \frac{B}{1\,{\rm G}} (1+z) \right]^{1/5} &\gtrsim& 12.5 \left( \frac{\nu_t}{1\,{\rm GHz}} \right) \left( \frac{S_t}{100\,{\rm mJy}} \right)^{-\frac{2}{5}} \left(\frac{\theta_t}{100\,{\rm mas}} \right)^{\frac{4}{5}}
\nonumber \\
\end{eqnarray}
Where the implied magnetic field estimate is sufficiently high, synchrotron emission could be argued to be physically implausible if it cannot provide a self-consistent explanation of the SED below the spectral turnover.  

The value of IPS-derived size measurements in informing the source physics can be illustrated with a hypothetical example. Suppose we observe a source with a spectral turnover frequency of $\nu_t=250\,$MHz at a flux density of $S_t=0.4\,$Jy, and that IPS is not observed in this source and thus yields a lower limit of $\theta = 1^{\prime \prime}$ on the source size at 150\,MHz.  Assuming that the turnover is due to synchrotron self-absorption and applying the lower limit on the source size, scaled to 250\,MHz assuming the standard $\theta \propto \lambda$ dependence for an optically thick synchrotron source, implies a lower limit to the source size at turnover of $\theta_t = 0.6^{\prime \prime}$, and hence a constraint on $B (1+z)$ of
\begin{eqnarray}
\left[ B (1+z) \right]^{1/5} &\gtrsim& 19.7 {\rm G}^{1/5}.
\end{eqnarray}
It could be argued that such a magnetic field is implausibly high.
For such a source at $z \approx 0$, the implied cyclotron frequency of $\Omega_e \sim 9 \times 10^{5}\,$GHz, and the fact that the synchrotron emission from an electron with Lorentz factor $\gamma$ peaks around a frequency $\omega_c = (3/2) \Omega_e \gamma^2$, leads to a contradiction, since it would imply that the hypothesized synchrotron emission would actually emanate from non-relativistic electrons for all $\nu \lesssim 10^{6}\,$GHz.

\subsection{High redshift source candidates}
The use of a steep spectral index to filter out high redshift radio galaxies in low radio frequency surveys \citep[eg.][]{McCarthy1993} showed great promise and most of the known high redshift radio galaxies have been found using this technique \citep{miley2008}.  However the highest redshift radio galaxy, TN J0924-2201 at z=5.2 \citep{vanBreugel1999} was found almost two decades ago and the dream of using high redshift radio galaxies to probe the early universe has not been fulfilled.  However using wide field low frequency IPS observations to determine the flux density in compact (kpc scale) components in combination with the spectral index may provide better constraints on the selection of galaxies at high redshift, minimizing the follow-up time need on large optical/IR telescopes. There are two mechanisms which generate redshift dependent effects we can exploit:
\begin{enumerate}
\item Synchrotron self-absorption at low rest frame frequencies must occur in some compact sources and will cause a spectral flattening unless the redshift is high enough to move the turnover to below 72\,MHz in the observed frame (Section \ref{Sec:PeakedSpectrumSources}). Hence the small number of sources in Figure \ref{Fig:avg-nonscintillatingAll} with steep spectral index but also high NSI (all the source is compact) could be at high redshift. 
\item The moderately scintillating sources (NSI $\sim$ 0.5) with integrated spectral index too steep to be flat spectrum core in steep spectrum radio lobes (Section \ref{Sec:PartiallyScintillatingSources} and Figure \ref{Fig:avg-nonscintillatingAll}) so they are likely to be hot spots which are a significant fraction of the total radio luminosity. This is more likely to be seen at high redshift where the diffuse lobes have been quenched by the high inverse Compton losses.
\end{enumerate}

\section{Interesting and unusual sources}
\label{sec:interesting}

\subsection{NGC 253}
\label{sec:ngc253}
NGC\,253 is one of the brightest starburst galaxies in the Southern Hemisphere.  Many supernovae remnants have been detected in the starburst region and it has a \fermi\ detection. We detect a weak (20\,mJy) scintillating component in our variability image. However, close inspection shows that the scintillating component is displaced 83\,arcsec (1.4\,kpc) from the nucleus and lies just outside the strong starburst region. We estimate a linear size limit of $<$ 5\,pc. \cite{UlvestadAntonucci1997} have obtained detailed VLA images at high radio frequencies which detect many individual compact sources in the starburst region. \cite{Lenc2006} and \cite{Rampadarath2014} have made VLBI observations at 2.3 GHz in the starburst region and show that strong free-free absorption is present which  will limit any detections of compact sources in the central starburst at frequencies below 2\,GHz. The position of the scintillating component is RA 00:47:27, Dec -25:17:46 with an error of 17\,arcsec. This is outside any of the existing high angular resolution images but the scintillating component may still be a supernovae remnant far enough away from the nucleus of NGC\,253 to avoid free-free absorption.

Supernova 1940E was discovered by Fritz Zwicky in NGC\,253 with an offset of 50\arcsec West from the centre \cite{Zwicky1948IAUC}, with the position of RA (J2000) 00:47:29.36 and Dec (J2000) -25:17:35.6 \citep{Lennarz2012A&A...538A.120L}, and is the only optically identified SNR in NGC 253. This position is within 2 sigma errors of our position. Thus, it is possible that we have detected the radio counterpart of SNR 1940E. High resolution continuum observations will be made in the future to confirm this identification.

\subsection{The extreme peaked \fermi\ source GLEAM J013243-165444}
\label{Sec:J013243-16544}
Of the \fermi\ sources exhibiting interplanetary scintillation, the radio SED of GLEAM J013243-16544 is conspicuous for its strong curvature across the MWA band.  The spectrum exhibits a concave shape peaked at $\approx 150\,$MHz (see Figure\,\ref{Fig:ips_Fermi}).  Moreover, the GLEAM spectral measurements are marginally inconsistent with the TGSS measurement at the corresponding frequency, and the spectral downturn evident in the MWA data appears inconsistent with its extrapolation to the MRC, and higher-frequency measurements.

While spectral curvature of the sort observed in GLEAM J013243-16544 is evident in several other GLEAM sources (see \cite{Callinghametal17}), it is pertinent to consider to what extent the spectral shape and the disparity between non-contemporaneous measurements of the SED of this source might be influenced by refractive {\it interstellar} scintillation (RISS). Such variability could explain the disparity between the the GLEAM spectrum and the TGSS measurement, taken 3 years earlier than the MWA spectrum, and the MRC data point, which pre-dates the MWA data by several decades \citep{Large1981}.
While it is also conceivable that the disparity is due to difference of the two flux density scales, we remark that the compactness of the source renders the effects of RISS likely for this source; the presence of IPS already establishes the source to be more compact than $\sim 0.3^{\prime \prime}$ and 6-km visibility \citep{Chhetri2013} from AT20G survey suggests its size to be more compact than $\sim 0.1^{\prime \prime}$ at 20\,GHz.  

RISS induces variability on a typical timescale of order years at metre-wavelengths, with intensity fluctuations that are correlated over a frequency range $\nu_{\rm dc}$ comparable to the observing frequency itself (i.e. $\nu_{\rm dc} \sim \nu$).   At the high Galactic latitude of this source, $b=-79^\circ$, the transition between weak and strong scintillation is predicted to occur at $\approx 7\,$GHz (see \cite{Walker1998}; \cite{Walker2001}; \cite{CordesLazio2002}), and the effective distance to the scattering material is $\sim 1\,$kpc.  We can estimate a characteristic timescale and refractive modulation index based on these fiducial numbers.  The characteristic timescale of the fluctuations is then
\begin{eqnarray}
t_{\rm RISS} = 2.4 \times 10^2 \,\left( \frac{v_{\rm ISS}}{30\,{\rm km\,s}^{-1}} \right)^{-1} \left( \frac{\nu}{150\,{\rm MHz}}  \right)^{-2} \,\, \hbox{days},
\end{eqnarray}
where the scintillation speed $v_{\rm ISS}$ is normalised to a value characteristic of the Solar System relative to the local standard of rest.  The modulation index corresponding the scintillations is 
\begin{eqnarray}
m_{\rm RISS} = 0.14 \left( \frac{\nu}{150\,{\rm MHz}}\right)^{17/30},
\end{eqnarray}
where the index $17/30$ follow from the assumption that the interstellar density inhomogeneities follow a Kolmogorov power spectrum.  The typical peak-to-peak amplitude of the variability is $3 m_{\rm ISS}$, indicating that variations of order 40\% of the mean flux density might be observed.  The foregoing timescale and amplitude estimates are valid for a source whose angular size does not exceed the refractive angular scale of 
\begin{eqnarray}
\theta_{\rm RISS} = 4.1 \left( \frac{\nu}{150\,{\rm MHz}}  \right)^{-2} \,\,\hbox{mas}
\end{eqnarray}
An intrinsic source size $\theta_{\rm src}$ greater than $\theta_{\rm RISS}$ suppresses the modulation index by a factor $\sim \theta_{\rm src}/\theta_{\rm RISS}$ and increases its variability timescale by a similar factor.  

We remark that the time span between the disparate observations of GLEAM J013243-16544 considerably exceeds the expected variability timescale, and that the amplitude of the deviations is comparable to the value expected of RISS. Thus is it possible that RISS is either wholly (or partially) responsible for the time-variable broadband deviations in the spectrum provided that the source size is less than (comparable to) $\sim 4\,$mas, corresponding to a source brightness temperature of $1.2 \times 10^{12}\,$K at 150\,MHz. Such a compact source size may be regarded as a prediction of this hypothesis.  

\subsection{Extreme steep spectrum sources} \label{Sec:ExtremeSteepSpecSrc}
GLEAM sources J010247-215651 and J004130-092221 in our catalogue show extreme steep spectral indices, with spectral indices of -2.36 and -2.31 respectively. Both objects are in our high S/N subsample. With NSI limits of 0.25 and 0.09 respectively, both objects are weak/non-scintillators. Source J004130-092221 has a single counterpart in NVSS and it maintains a very steep spectrum of -2.40 between 162 and 1400 MHz. Source J010247-215651 does not have a NVSS counterpart in a 60 arcsecond radius search. 

Both these objects are known cluster halo/relics associated with clusters A85 and A133 respectively, and MWA observations of these and other similar objects are discussed in detail in Duchesne et al. (submitted) and Johnston-Hollitt et al. (in prep). 

Any very steep spectrum source with low NSI can, thus, be eliminated from searches for pulsars where the candidates are drawn based on steep spectra alone, but they can be tracers of clusters. Alternatively, if very steep spectra objects show low to moderate scintillation, such objects become candidate radio galaxies exhibiting radiation loss (relic galaxies) but with possible regenerated activity in the core. Very steep spectrum objects with strong scintillation, on the other hand, are excellent pulsars candidates, as discussed below.

\subsection{The Pulsar PSR J0034-0721 and pulsar searching}
\label{sec:pulsar}
Searching for new pulsars is one of the key science drivers for the international SKA project.
The traditional time-based searches at full sensitivity require phased array beams which optimize sensitivity but make it difficult to cover large areas of the sky.
An alternative approach \citep[e.g.][]{Dai2016} is to search in continuum images for sources with small enough angular size to scintillate.
Only pulsars are compact enough to exhibit diffractive interstellar scintillation (DISS).
However searches that rely on this effect to detect pulsars are necessarily limited by the fact that the characteristic bandwidth and timescale of DISS decreases sharply with distance.
The requirement that the scintillations be resolvable in both frequency and time therefore strongly constrains the detection volume of DISS-based pulsar surveys.

All pulsars will have strong IPS independent of their location in the Galaxy.
The problem then becomes the contamination from other sources (AGN) which may also have small enough diameter to have strong IPS.
By combining the requirement of strong IPS (NSI $>$ 0.9) and a steep spectral index (alpha $<$ -0.7) we reduce the number of contaminating AGN in our sample by a factor of $\sim$ 45 making conventional pulse-searching follow-up much more practical.

There are 10 known pulsars within this high galactic latitude field (in the ATNF pulsar catalogue\footnote{http://www.atnf.csiro.au/people/pulsar/psrcat/}) but based on the extrapolation of their measured flux densities at 400 and 1400 MHz, most are far too weak to be detected in this observation. 
However, as noted in section~\ref{sec:var_image}, one pulsar was detected at $>$5$\sigma$ in variability image, and was only just below our 5$\sigma$ cut in the standard image.
The scintillation index and NSI are 0.64$\pm$0.16 and 1.92$\pm$0.49 respectively, the latter being higher than for any source in our catalogue.

The pulsar has also been detected as an MWA continuum source in GLEAM data by \cite{Murphy2017b}, who measured a flux density of $292\pm14$\,mJy at 200\,MHz, and appears in the GLEAM extragalactic catalogue as GLEAM\,J003408-072153.
\cite{Bell2016} found that this pulsar was a variable continuum source at 154\,MHz and concluded that refractive scintillation was the most plausible cause of the variability they observed. 

This detection of a known and relatively strong pulsar shows that an MWA IPS observation with an RMS of $\sim$100\,mJy (see paper I) is on the threshold of the sensitivity needed to make new pulsar detections.  It is not very efficient to use longer integration times since sensitivity only improves as the fourth root of the observing time in the low S/N limit (see equation 5 in paper I), however the sensitivity will be dramatically improved using future telescopes with higher instantaneous sensitivity.  For example scaling the MWA sensitivity to the proposed SKA1-low collecting area and bandwidth gives a factor of 300 improvement so a 6 min IPS observation with SKA1-low would reach 340\,$\mu$Jy at 160\,MHz corresponding to 10\,$\mu$Jy at 1.4GHz for a typical pulsar spectrum.  This sensitivity approaches the desired SKA blind pulsar survey requirements but with a larger FoV and simpler backend.

A further possibility is to look for any unassociated gamma-ray sources with steep spectrum scintillating counterparts since they will be very good pulsar candidates.

\section{Summary}
\label{sec:summary}
In the era of new sensitive telescopes, where angular resolution and the wide field-of-view have not yet found mutual inclusiveness for routine observations, we find that IPS provides a powerful alternative to identify subarcsecond compact structures and to determine their astrophysical properties. We have identified compact structures in 302 objects out of a total of 2550 objects in the 30$\times$30 sq deg. field of observation made with the MWA, using a $<$ 5 minute of observation at 162 MHz.  This work has shown that our time domain technique is both powerful and extremely efficient compared to conventional techniques. The angular sizes show excellent consistency with other surveys (e.g. FIRST, TGSS). Our future wider survey will form an excellent high angular resolution catalogue to complement all-sky surveys at low frequencies.

Using a high signal to noise ratio subset, we find that almost half of the total population show evidence of subarcsecond compact structures, at 0.3 arcsec scales, and $\sim$ 10 \% of compact sources are not associated with any extended structures. In comparison, the high frequency AT20G survey has $\sim$80\% compact objects at sub arcsecond scales while the NVSS survey has $\sim$ 25 \%. Thus, this fits in well with the picture where the fraction of compact source population is becoming smaller with increasing wavelength. Interestingly, we find that the low frequency subarcsecond compact population is dominated by peaked spectrum objects that are relatively rare at higher frequencies. While beyond the scope of this paper, the angular size constraints placed by our technique on these peaked spectrum objects can be used to constrain the mechanism that gives rise to their spectra and also the evolution of their peak frequencies.

This work outlines two possibilities for identifying high redshift radio galaxies:
\begin{enumerate}
\item steep-spectrum compact objects (strongly scintillating objects) with their synchrotron self-absorbed peaks shifted below GLEAM frequencies due to redshift.
\item hotspots embedded in extended structures (moderately scintillating objects) with steep integrated spectral index, where the hotspots provide a significant fraction of total luminosity, seen at high redshift due to the diffuse lobes being quenched by high inverse Compton losses.
\end{enumerate}
Our future work will investigate this new possibility of identifying high redshift objects from low frequencies.

We find no correlation between sources that show gamma ray emission (from \fermi) and their angular size estimates from scintillation indices. This is in sharp contrast to the results from high frequencies where counterparts of \fermi\ sources (at 20 GHz) are almost entirely subarcsecond compact objects. Considering we find that the compact population at 162 MHz is dominated by peaked spectrum objects, we posit that at low frequencies we are observing an emergence of a different compact source population from those at gigahertz frequencies.

We find that $\sim$ 3 \% of strongly scintillating objects to show greater than 10 percent variability in the timescales of 1-3 years. Thus, compact objects identified in this (and subsequent) publication can be expected to be part of calibrators for the current low frequency radio telescopes and the future SKA-low. 

Our detection of a known pulsar, PSR J0034-0721 demonstrates that this image plane technique is at the threshold of sensitivity required to detect new pulsars. Future low frequency radio telescopes, such as the SKA-low, will have sufficient sensitivity on the IPS timescale to detect a new population of weaker pulsar candidates over the full field of view. This IPS filter will greatly reduce the number of phased array beams needed for timing follow-up. 

Future observations of IPS with the MWA will take two paths:  
i) multiple observations of the same field multiple at different solar elongation and different solar latitude. This has the double advantage of averaging out any variability in the solar wind conditions, and measuring the IPS at smaller elongations and hence with greater sensitivity. \newline
ii) observations of new fields across the sky that are within $\sim$50 degrees of the ecliptic.  It will be easy to obtain these short observations in the daytime and many more fields have already been observed. The limitations will be almost entirely due to the computing resources needed.

These future observations will provide an even greater coverage of the sky, enabling further investigation of AGN populations with better statistics, possible discovery of new pulsars, candidate calibrators for future low frequency instruments and potentially the discovery of new rare classes of radio sources.  Further refinement in IPS techniques can be expected to provide new and innovative ways to explore the high angular resolution parameter space with SKA-low unshackled from the ionospheric limitations.

\section*{Acknowledgments} 
We thank D. Kaplan for insightful comments on an early draft of this paper. R.C. also thanks  T. Franzen and M. Johnston-Hollitt for very useful discussions. R.C. thanks C. Hunt at Carnegie Observatories for her assistance in finding IAU circular 848, and the scientific editor for facilitating this.
This scientific work makes use of the Murchison Radio-astronomy Observatory, operated by CSIRO. We acknowledge the Wajarri Yamatji people as the traditional owners of the Observatory site. Support for the operation of the MWA is provided by the Australian Government (NCRIS), under a contract to Curtin University administered by Astronomy Australia Limited. We acknowledge the Pawsey Supercomputing Centre which is supported by the Western Australian and Australian Governments.
Parts of this research were conducted by the Australian Research Council Centre of Excellence for All-sky Astrophysics (CAASTRO), through project number CE110001020. 
We used the TOPCAT software \citep[]{Taylor2005} for the analysis of some of our data.




\bibliographystyle{mnras}
\bibliography{references-IPS-SingleField}







\bsp	
\label{lastpage}
\end{document}